\documentclass[
aps,%
11pt,%
final,%
notitlepage,%
oneside,%
twocolumn,%
nobibnotes,%
nofootinbib,%
superscriptaddress,%
noshowpacs,%
centertags]%
{revtex4}

\def\saoname{Special Astrophysical Observatory of the Russian AS,
              Nizhnij Arkhyz 369167, Russia}

\def\squareforqed{\hbox{\rlap{$\sqcap$}$\sqcup$}}

\def\sq{\ifmmode\squareforqed\else{\unskip\nobreak\hfil
\penalty50\hskip1em\null\nobreak\hfil\squareforqed
\parfillskip=0pt\finalhyphendemerits=0\endgraf}\fi}

\def\degr{\hbox{$^\circ$}}

\def\arcsec{\hbox{$^{\prime\prime}$}}

\def\utw{\smash{\rlap{\lower5pt\hbox{$\sim$}}}}

\def\udtw{\smash{\rlap{\lower6pt\hbox{$\approx$}}}}

\def\fm{\hbox{$.\!\!^{\rm m}$}}

\def\farcm{\hbox{$.\mkern-4mu^\prime$}}

\def\diameter{{\ifmmode\mathchoice
{\ooalign{\hfil\hbox{$\displaystyle/$}\hfil\crcr
{\hbox{$\displaystyle\mathchar"20D$}}}}
{\ooalign{\hfil\hbox{$\textstyle/$}\hfil\crcr
{\hbox{$\textstyle\mathchar"20D$}}}}
{\ooalign{\hfil\hbox{$\scriptstyle/$}\hfil\crcr
{\hbox{$\scriptstyle\mathchar"20D$}}}}
{\ooalign{\hfil\hbox{$\scriptscriptstyle/$}\hfil\crcr
{\hbox{$\scriptscriptstyle\mathchar"20D$}}}}
\else{\ooalign{\hfil/\hfil\crcr\mathhexbox20D}}%
\fi}}

\newcommand{\ab}{Astrophysical Bulletin}

\newcommand{\aj}{Astronom. J.}
\newcommand{\apjs}{Astrophys. J. Suppl.}

\newcommand{\araa}{Annu. Rev. Astronom. Astrophys.}

\newcommand{\bsao}{Bull. Spec. Astrophys. Obs.}

\newcommand{\mnras}{Monthly Notices Roy. Astronom. Soc.}

\begin{document}
\selectlanguage{english}




\title{Ultra-Flat Galaxies Selected from RFGC Catalog. III. Star Formation Rate}

\author{\firstname{O.~V.}~\surname{Melnyk}}
 \email{melnykol@gmail.com}
 \affiliation{University of Zagreb,  Zagreb, 10000 Croatia}

\author{\firstname{V.~E.}~\surname{Karachentseva}}
\affiliation{Main Astronomical Observatory of National Academy of Sciences of Ukraine, Kiev, 03143 Ukraine}

\author{\firstname{I.~D.}~\surname{Karachentsev}}
\email{ikar@sao.ru} \affiliation{\saoname}

\received{December 21, 2016} \revised{January 03, 2017}

\begin{abstract}
We examine the star formation properties of galaxies with very
thin disks selected from the Revised Flat Galaxy Catalog (RFGC).
The sample contains 333 ultra-flat galaxies (UFG) at high Galactic
latitudes, $|b|>10\degr$, with a blue major angular diameter of $a
\geq 1\farcm2$, blue and red apparent axial ratios of $(a/b)_b >
10$, $(a/b)_r > 8.5$ and radial velocities within
10\,000~km\,s$^{-1}$. As a control sample for them we use a
population of 722 more thick RFGC galaxies with $(a/b)_b > 7$,
situated in the same volume. The UFG distribution over the sky
indicates them as a population of quite isolated galaxies. We
found that the specific star formation rate, $sSFR_{\rm FUV}$,
determined via the FUV GALEX flux, increases steadily from the early type
 to  late type disks for both the UFG and RFGC--UFG samples,
showing no significant mutual difference within each morphological
type $T$. The population of UFG disks has the average
H\,I-mass-to-stellar-mass ratio  by $(0.25\pm0.03)$~dex higher than
that of  RFGC--UFG galaxies. Being compared with arbitrary
orientated disks of the same type, the ultra-flat edge-on galaxies
reveal that their total H\,I mass is hidden by self-absorption on
the average by approximately 0.20~dex. We demonstrate that using the
robust stellar mass estimate via $\langle B-K \rangle$-color and
galaxy type $T$ for the thin disks, together with a nowaday
accounting for internal extinction, yields their $sSFR$ quantities
definitely lying below the limit of $-9.4$~dex\,(yr$^{-1}$). The
collected observational data on UFG disks imply that their average
star formation rate in the past has been approximately three times
the current $SFR$.  The UFG galaxies have also sufficient amount
of gas to support their observed $SFR$ over the following nearly
9~Gyrs.

\end{abstract}

\maketitle

\section{INTRODUCTION}

Since the time of the first studies~\cite{tin1968:Melnyk_n_en,
lar1980:Melnyk_n_en}, star formation in  galaxies is examined at the
scales of the Local Volume (at the distances up to
10~Mpc~\cite{ken2008:Melnyk_n_en,ken2009:Melnyk_n_en,lee2011:Melnyk_n_en}),
of the nearby universe (approximately up to
50~Mpc~\cite{thi2007:Melnyk_n_en,bua2007:Melnyk_n_en}),  as well as
at $z\sim 1$, where the evolutional effects become
significant~\cite{bri2004:Melnyk_n_en,noe2007:Melnyk_n_en,abr2014:Melnyk_n_en}.
The methods determining the star formation rate (SFR) from radiation
in the far and near ultraviolet ranges, in the H$\alpha$ line, from
the equivalent widths of  spectral lines, etc. were developed (see
the survey~\cite{ken1998:Melnyk_n_en}). Detailed optical and radio
observations were conducted and SFR was determined  for small samples
of nearby galaxies
(e.g.,~\cite{god2004:Melnyk_n_en,kar2015:Melnyk_n_en}).

A massive study of star formation in the galaxies became possible
owing to the emergence of large-scale sky surveys, carried out in
ultraviolet, optical and infrared ranges: GALEX~\cite{mar2005:Melnyk_n_en}, SDSS~\cite{aba2003:Melnyk_n_en}, 2MASS~\cite{scr2006:Melnyk_n_en,jar2000:Melnyk_n_en}, WISE~\cite{jar2012:Melnyk_n_en}. Advantages and
 disadvantages of determining the SFR  in the most nearby  (the possibility of
individual approach) and distant galaxies (sample sizes of
about  tens of thousands of objects) are obvious.

A general conclusion from a vast array of publications (without
touching the farthest objects) is that star formation in elliptical
galaxies, not containing gas and dust, is quenched, while spiral and
irregular dwarf galaxies demonstrate a high star formation rate,
sufficient to explain the observed amount of their stellar mass.

For the galaxies of the nearby universe within \mbox{$D\sim50$~Mpc}
the upper limit of specific star formation rate was found,  $\log
sSFR
=-9.4$~yr$^{-1}$~\cite{kar2013a:Melnyk_n_en,kar2013b:Melnyk_n_en,mel2015:Melnyk_n_en}.
 This condition is met for the Local Volume galaxies   ($D<10$~Mpc),
located in the environments of different
density~\cite{kar2013a:Melnyk_n_en}, as well as for the
LOG~\cite{kar2011:Melnyk_n_en}  and 2MIG~\cite{kara2010:Melnyk_n_en}
catalogs of isolated galaxies, having the depth of   $z~\sim~0.01$.

The studies devoted to star formation in spiral galaxies usually
consider the objects arbitrarily inclined to the line of sight~\cite{mor2016:Melnyk_n_en}.
Only a few authors have separately studied the highly inclined (edge-on) spirals
(see~\cite{mat2000:Melnyk_n_en} and the references therein).

Heidmann et al.~\cite{hei2000:Melnyk_n_en} have earlier found that
namely the late-type spirals  are characterized by the highest axial
ratios $a/b$, where  $a$ and $b$ are the major and minor angular
diameters. As it was shown by
Karachentsev~\cite{kar1989:Melnyk_n_en}, selection of galaxies based
on their  apparent axial ratio  $a/b\geq7$ allows to choose flat
disks without noticeable signs of the bulge from an array of spirals.
The  application of this simple selection criterion led to the
creation of flat galaxy catalogs: Flat Galaxy Catalog
(FGC~\cite{kar1993:Melnyk_n_en}) and its improved version, the
Revised Flat Galaxy Catalog (RFGC~\cite{kar1999:Melnyk_n_en})
covering the entire sky.

Apart from the morphological features, flat late-type spirals  have
other peculiarities. Firstly, they are  distributed over the sky quite
homogeneously, without any visible clumps in the areas occupied by
clusters of galaxies~\cite{kar1993:Melnyk_n_en}. A more detailed examination
has shown that about 60\% of ultra-flat edge-on galaxies are isolated, while about 30\% belong to diffuse associations of galaxies, and only 10\% have nearby,  physically bound neighbors~\mbox{\cite{kara2016:Melnyk_n_en,kar2016:Melnyk_n_en}}. It can therefore be
assumed that star formation in these objects is mainly due to the
 internal processes, not being affected by nearby
neighbors. Secondly,  late spirals are characterized by
a high degree of detection in the H\,I~21~cm line. This allows
 to determine  their maximum amplitude of rotation
$V_m$ with a good accuracy, which is used in particular to account
for the internal extinction of light.

Note that the systematic study of star formation in the RFGC spiral
galaxies  was made possible with the advent of the GALEX ultraviolet
sky survey~\cite{mar2005:Melnyk_n_en}.  The purpose of the present study is to
compare the SFR in  ultra-flat  and flat edge-on galaxies. Both
samples were taken from the RFGC catalog. Special attention is given
to different methods of estimation of the internal extinction $A^i$
and determining the integral stellar mass of the galaxy $M_*$.

Section~2 describes the samples of flat and ultra-flat galaxies.
Section~3 discusses the ways of accounting for the internal
extinction in  edge-on galaxies. Section~4 briefly describes the
features of determination of the FUV flux and SFR. Section~5 is
devoted to different ways of estimation of the stellar mass of
galaxies, and selecting the most optimal of them. Section~6 presents
our main findings. Discussion and conclusions are given in Section~7.

\section{ORIGINAL DATA FOR ULTRA-FLAT GALAXIES}

The RFGC catalog~\cite{kar1999:Melnyk_n_en} contains  $N= 4236$ galaxies with
 angular diameters   $a\geq0\farcm6$ and apparent axial ratios
$a/b\geq7$ on the blue images of the  First Palomar Sky survey. From
the RFGC catalog we have selected ultra-flat galaxies  (UFG,
$N=817$),  for which the mutual conditions for the ``blue'' and
``red'' axial ratios were met: $(a/b)_{\rm blue}\geq 10$, $(a/b)_{\rm
red}\geq8.5$.  To reduce the impact of the selection effects, we
added  restrictions on the angular  diameter, Galactic latitude and
radial velocity:
 $a_{\rm blue}\geq1\farcm0$, $|b|>10\degr$, $V_h<10\,000$~km\,s$^{-1}$.
The number of thin UF galaxies amounted to $N=441$.

The application of the Schmidt test has shown that the completeness
of the RFGC catalog at \mbox{$a_{\rm blue}\geq1\farcm0$} is 70\%, and
the completeness of about 90\% is achieved at the   angular diameter
of \linebreak \mbox{$a_{\rm blue}=1\farcm2$}~\cite{kar1999:Melnyk_n_en}. We have
strengthened the selection criteria to ensure the acceptable
completeness (90\%), reduce the selection effects, and at the same
time to have a rather large number of galaxies remaining. Flat
galaxies with the following characteristics were selected from the
RFGC catalog:` $a_{\rm blue}\geq1\farcm2$, $(a/b)_{\rm blue}\geq7$,
$|b|>10\degr$, $V_h<10\,000$~km\,s$^{-1}$.

Some galaxies have been excluded from the consideration based on the following:
\begin{list}{}{
\setlength\leftmargin{2mm} \setlength\topsep{2mm}
\setlength\parsep{0mm} \setlength\itemsep{2mm} }
 \item 1) poor photometry due to a projected bright star (cases of
RFGC 514, 1543, 1747 and 3830);
 \item 2) a discrepancy between the direct distance estimate
  and  the distance determined from  radial velocity,
RFGC\,384, 2239 and 2246;
 \item 3) galaxy RFGC\,2614 was included in the sample by mistake (in the  LEDA database it is given  \linebreak $V_h =
11\,354$ km\,s$^{-1}$).
 \end{list}

Therefore, in our  truncated  RFGC catalog sample (further  in this
paper this sample is referred to as the  RFGC sample) $N=1055$  flat
galaxies are remaining. Among them, the population of ultra-flat
galaxies (UFG) numbers $N=333$ objects.

The following RFGC catalog~\cite{kar1999:Melnyk_n_en} data were used:
equatorial coordinates of  galaxies, their ``blue''
 and ``red'' axial ratios $a/b$, and morphological types.
 From the NED\footnote{\url{http://nedwww.ipac.caltech.edu}} and
HyperLEDA\footnote{\url{http://leda.univ-lyon1.fr}}~\cite{2014A&A...570A..13M:Melnyk_n_en} databases the heliocentric radial velocities $V_h$ were taken and converted to
 $V_{\rm LG}$, according to~\cite{kar1996:Melnyk_n_en}. From the HyperLEDA we have
also adopted the values of the apparent total magnitude $B_t$ (for
1055 RFGC objects and 333  UFG objects), the maximum rotation
velocity $V_m$ (979 and 311) and the magnitude in the 21~cm line
$m_{21}$ (831 and 258). The $g$, $r$, $i$ magnitudes at $0<g-r<1.1$
(448 and  156) were taken from the  SDSS (DR13) survey, while from
the 2MASS survey we took the $K_s$ values (837 and  235), and from
the  WISE---the magnitudes for 988 and 301 galaxies, respectively.
The magnitudes in the latter three surveys were determined in the AB
system.

\begin{figure*}[]
 \setcaptionmargin{5mm} \onelinecaptionsfalse \captionstyle{normal}
\includegraphics[width=\textwidth]{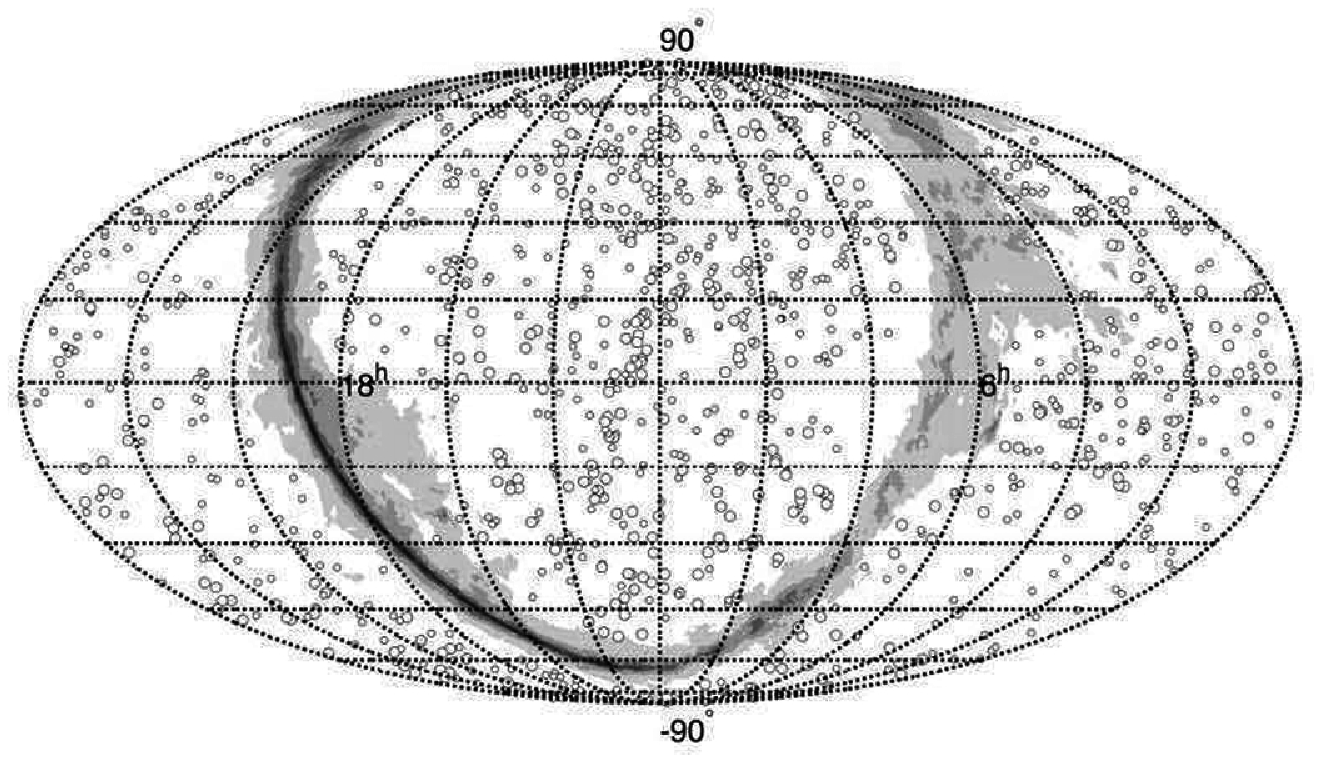}
\includegraphics[width=\textwidth]{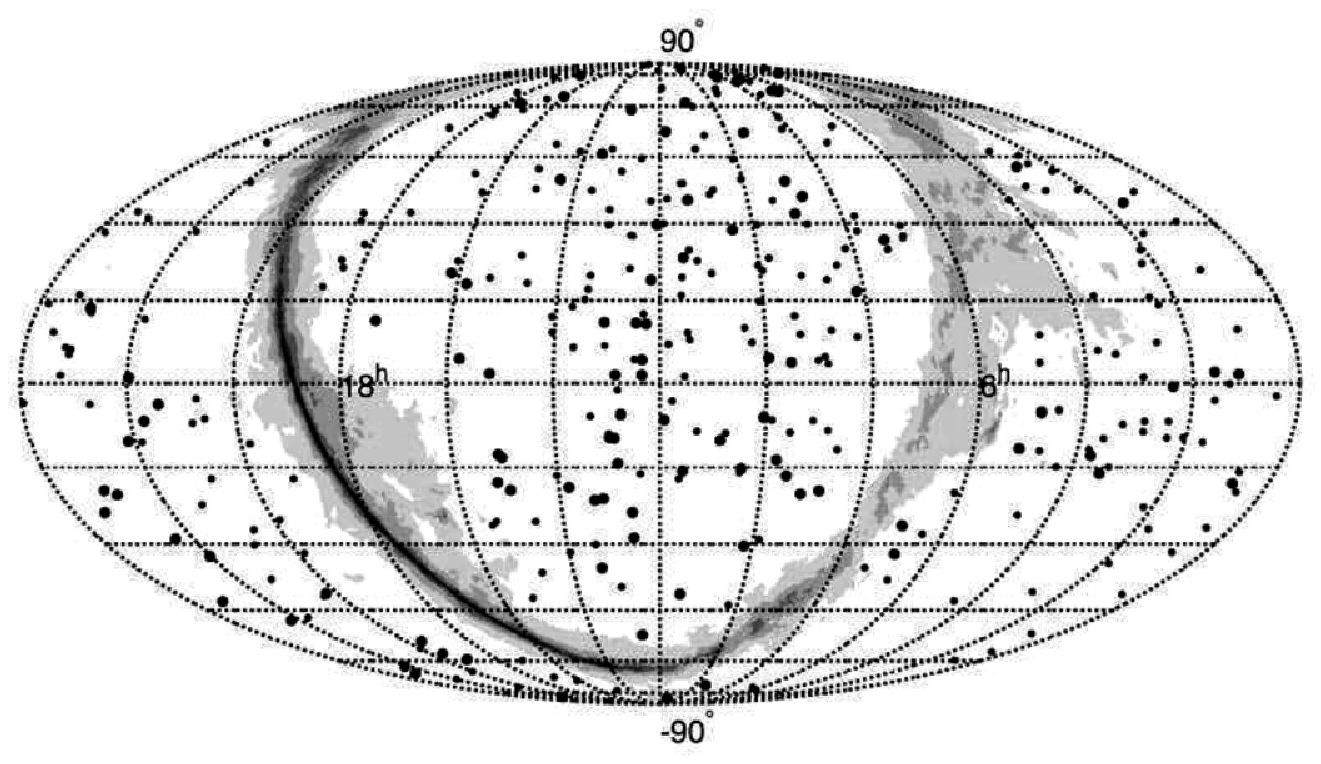}
\caption{The sky distribution in equatorial coordinates of the flat
galaxies RFGC, $N = 1055$ (empty circles), and  ultra-flat UFG, $N =
333$ (filled circles).  Gray fill marks the area of strong Galactic
extinction  $|b| < 10\degr$. Large circles mark the galaxies with
$V_{\rm LG} <3500$~km\,s$^{-1}$, and small circles---with
\mbox{$V_{\rm LG} = 3500-10\,000$}~km\,s$^{-1}$.}
\label{fig1:Melnyk_n_en}
\end{figure*}

Figure~\ref{fig1:Melnyk_n_en} shows the sky distribution in the
equatorial coordinates of the selected  flat   $N=1055$ (top panel)
and ultra-flat galaxies $N=333$ (bottom panel). Galaxies possessing
radial velocities \mbox{$V_{\rm LG} < 3500$}~km\,s$^{-1}$ are marked
with larger symbols. The region with the Galactic latitude  $|b| <
10\degr$ is painted gray. There is a weak concentration of nearby
flat galaxies along the equator of the Local Supercluster and in the
region of the nearby scattered cloud in Canes Venatici. The effect is
almost not noticeable for  ultra-flat galaxies, what
 indicates a higher fraction of isolated
galaxies among them.

 \begin{figure*}[]
 \setcaptionmargin{5mm} \onelinecaptionsfalse \captionstyle{normal}
\includegraphics[width=0.9\columnwidth]{Melnyk_fig2a.eps}
\includegraphics[width=0.9\columnwidth]{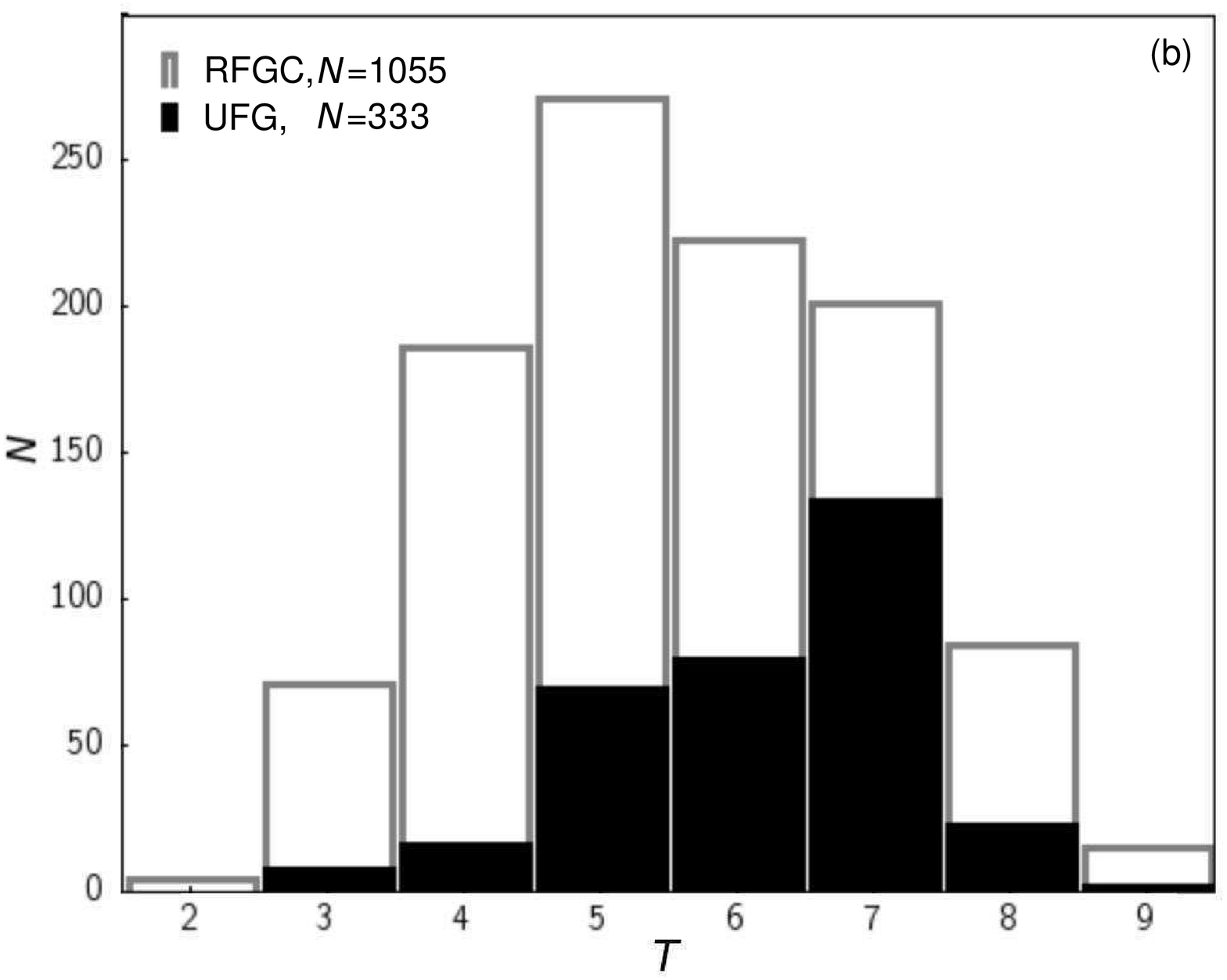}
\includegraphics[width=0.9\columnwidth]{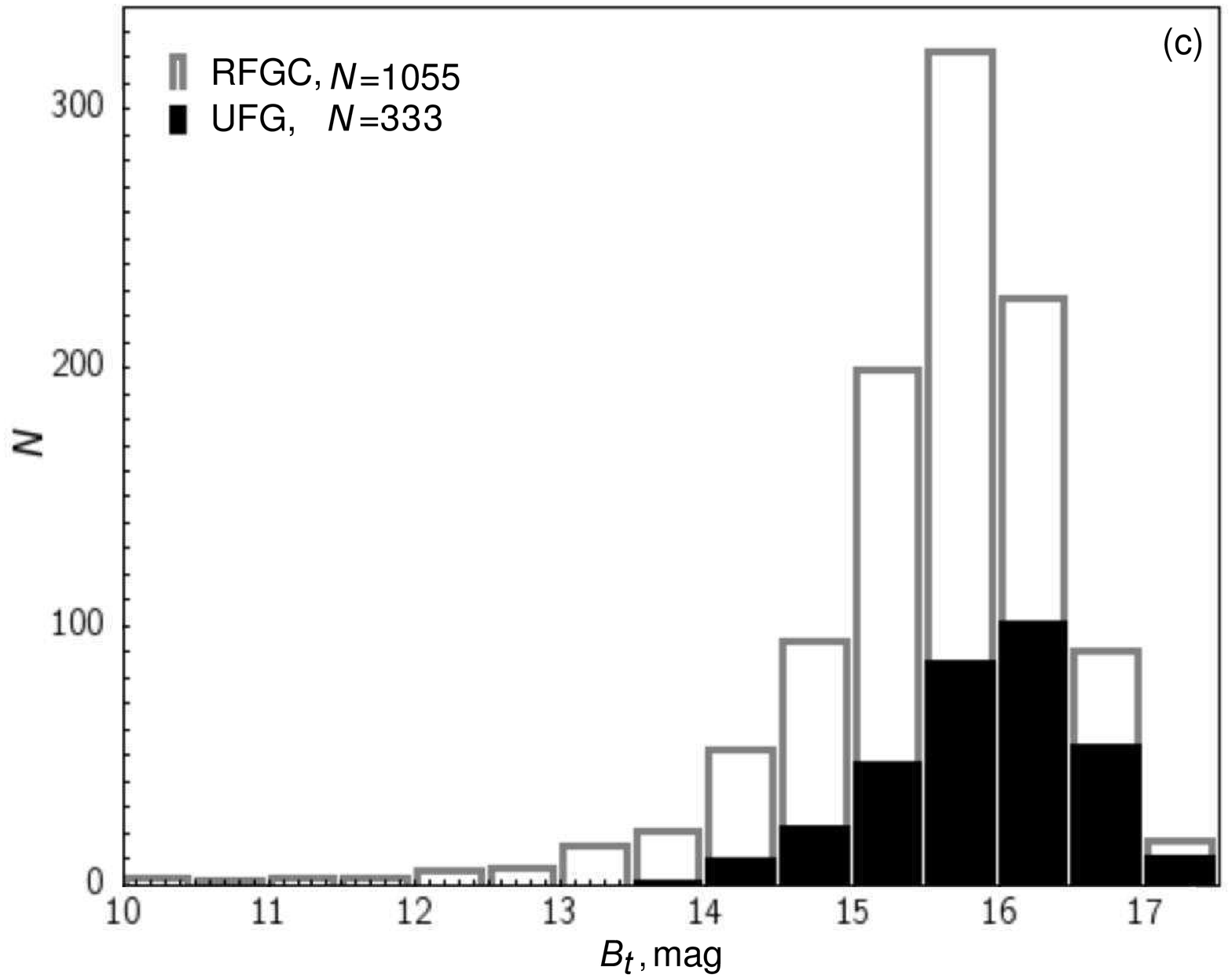}
\includegraphics[width=0.9\columnwidth]{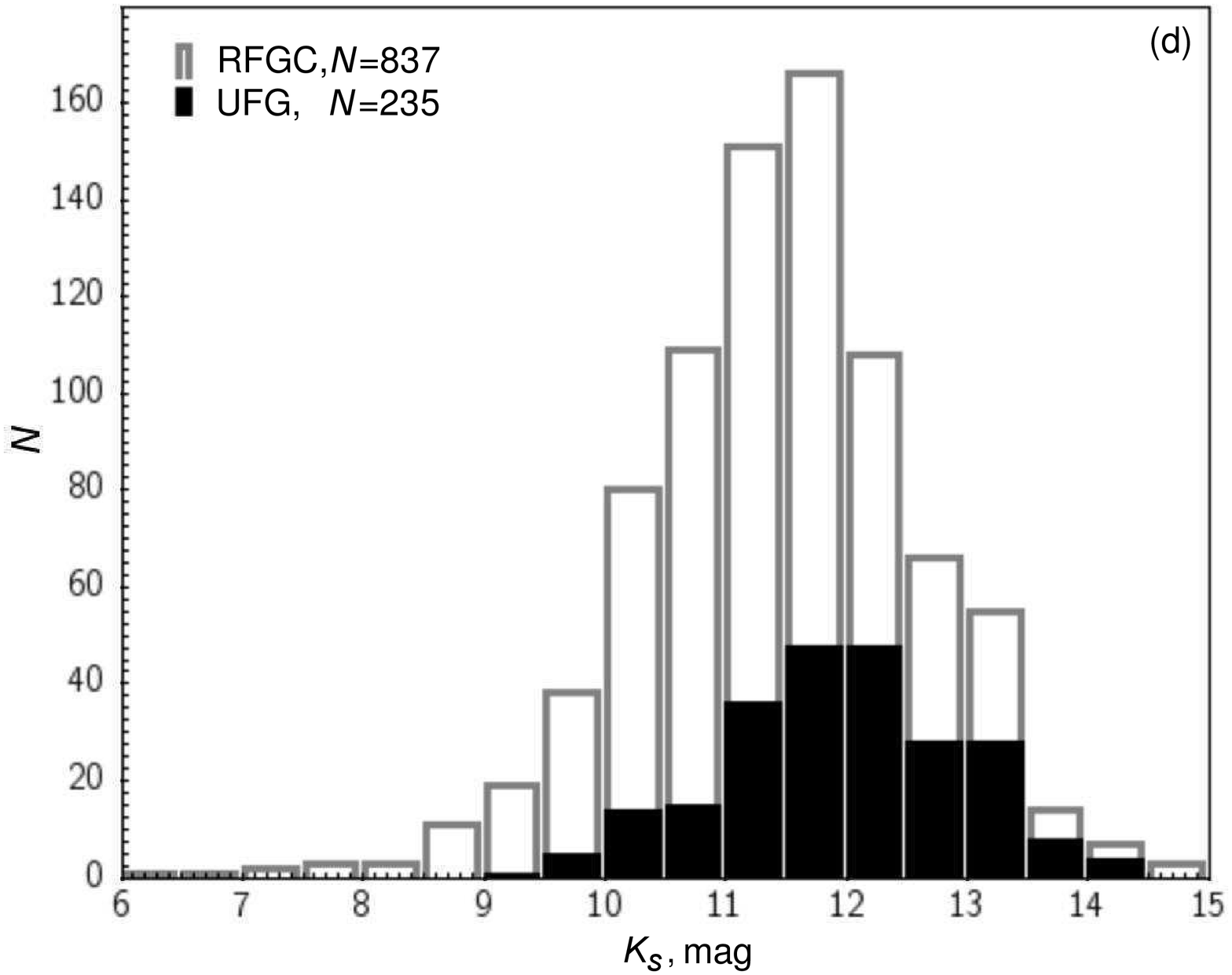}
 \caption{Histograms of the distribution of the number of galaxies based on  the observed
characteristics:
(a) radial velocity  $V_{\rm LG}$; (b)  morphological type $T$;
 (c)  the apparent total magnitude  $B_t$ from the  HyperLEDA database;
 (d)  the 2MASS survey   $K_s$ magnitude.
The data relating to ultra-flat galaxies, UFG, are darkened.}
 \label{fig2:Melnyk_n_en}
 \end{figure*}

Figure~\ref{fig2:Melnyk_n_en} presents a general understanding of the distribution of the main observed characteristics. On each panel there,
ultra-flat galaxies are marked with black color. As seen from the
distribution based on radial velocity, the sample of \mbox{UFG galaxies}
 has approximately the same depth as the sample of  RFGC galaxies.
Consequently, the effect of the galaxy selection  based on luminosity   does not play
a significant role in the comparison of properties of UFG and
RFGC populations.

The distribution of RFGC galaxies based on morphological types has a peak for the    Sc--Scd types ($T = 5$--$6$), while in the
UFG galaxies the peak is shifted to a later type Sd ($T = 7$).
This once again emphasizes the well-known fact that the thinnest
disks are typical of galaxies, the spiral structure of which is
 on the verge of transition from a regular to chaotic pattern.

As shown in Fig.~ 2c, the considered RFGC galaxy sample of galaxies has
a photometric completeness to the apparent magnitude of \mbox
{$B_t\simeq16\fm0$}. For the  UFG galaxies the completeness limit is
shifted by   \linebreak \mbox{$\Delta B\simeq+0\fm5$},
and the
systematic difference is preserved within each morphological type.

Figure~\ref{fig2:Melnyk_n_en}d indicates that in the photometric
$K_s$\mbox{-}band of the
 2MASS survey both samples start loosing  their completeness
at  \mbox{$K_s\simeq12\fm0$}. For many flat
galaxies having a bluish color and low surface brightness,
2MASS-photometry at  $K_s>12\fm0$  proves to be burdened with
significant systematic errors.

\begin{table*}
\setcaptionmargin{0mm} \onelinecaptionstrue \captionstyle{normal}
\caption{Average values and standard errors of the mean for the
original characteristics of the flat and ultra-flat galaxies
depending on the morphological type}
\medskip
\begin{tabular}{l|r@{$\,\pm\,$}l|r@{$\,\pm\,$}l|r@{$\,\pm\,$}l|r@{$\,\pm\,$}l|r@{$\,\pm\,$}l|r@{$\,\pm\,$}l|r@{$\,\pm\,$}l} \hline
  \multicolumn{1}{c|}{Parameter}  &\multicolumn{14}{c}{Morphological type}\\ \cline{2-15}
                & \multicolumn{2}{c|}{All types} & \multicolumn{2}{c|}{2\,+\,3} & \multicolumn{2}{c|}{4} & \multicolumn{2}{c|}{5} & \multicolumn{2}{c|}{6} & \multicolumn{2}{c|}{7} & \multicolumn{2}{c}{8\,+\,9} \\ \hline
$(a/b)_{b}$  & \multicolumn{2}{c|}{$N_{\rm RFGC-UFG}=722$} & \multicolumn{2}{c|}{67} & \multicolumn{2}{c|}{170} & \multicolumn{2}{c|}{201} & \multicolumn{2}{c|}{143} & \multicolumn{2}{c|}{67}  & \multicolumn{2}{c}{74} \\
             & 8.53 & 0.04                                 & 8.15 & 0.14             & 8.35 & 0.07              & 8.62 & 0.09              & 8.87 & 0.10              & 8.68 & 0.14              & 8.24 & 0.12 \\
             & \multicolumn{2}{l|}{$N_{\rm UFG}=333$}      & \multicolumn{2}{c|}{8}  & \multicolumn{2}{c|}{16}  & \multicolumn{2}{c|}{70}  & \multicolumn{2}{c|}{80}  & \multicolumn{2}{c|}{134} & \multicolumn{2}{c}{25} \\
             & 12.74 & 0.13                                & 10.33 & 0.08            & 11.60 & 0.33             & 12.32 & 0.23             & 12.51 & 0.24             & 13.50 & 0.24             & 12.04 & 0.35 \\ \hline
$(a/b)_{r}$  & \multicolumn{2}{c|}{$N_{\rm RFGC-UFG}=722$} & \multicolumn{2}{c|}{67} & \multicolumn{2}{c|}{170} & \multicolumn{2}{c|}{201} & \multicolumn{2}{c|}{143} & \multicolumn{2}{c|}{67}  & \multicolumn{2}{c}{74} \\
             & 7.54 & 0.04                                 & 6.91 & 0.14             & 7.21 & 0.07              & 7.69 & 0.07              & 8.01 & 0.12              & 7.72 & 0.15              & 7.34 & 0.12 \\
             & \multicolumn{2}{l|}{$N_{\rm UFG}=333$}      & \multicolumn{2}{c|}{8}  & \multicolumn{2}{c|}{16}  & \multicolumn{2}{c|}{70}  & \multicolumn{2}{c|}{80}  & \multicolumn{2}{c|}{134} & \multicolumn{2}{c}{25} \\
             & 11.27 & 0.11                                & 9.19 & 0.31             & 9.99 & 0.31              & 10.55 & 0.16             & 11.05 & 0.19             & 11.98 & 0.18             & 11.59 & 0.52 \\ \hline
$B_t$        & \multicolumn{2}{c|}{$N_{\rm RFGC-UFG}=722$} & \multicolumn{2}{c|}{67} & \multicolumn{2}{c|}{170} & \multicolumn{2}{c|}{201} & \multicolumn{2}{c|}{143} & \multicolumn{2}{c|}{67}  & \multicolumn{2}{c}{74} \\
             & 15.40 & 0.03                                & 15.22 & 0.11            & 15.39 & 0.07             & 15.48 & 0.06             & 15.32 & 0.10             & 15.45 & 0.13             & 15.49 & 0.11 \\
             & \multicolumn{2}{l|}{$N_{\rm UFG}=333$}      & \multicolumn{2}{c|}{8}  & \multicolumn{2}{c|}{16}  & \multicolumn{2}{c|}{70}  & \multicolumn{2}{c|}{80}  & \multicolumn{2}{c|}{134} & \multicolumn{2}{c}{25} \\
             & 15.92 & 0.04                                & 15.61 & 0.23            & 15.99 & 0.13             & 15.88 & 0.08             & 15.93 & 0.07             & 15.98 & 0.06             & 15.77 & 0.18 \\ \hline
$K_s$        & \multicolumn{2}{c|}{$N_{\rm RFGC-UFG}=602$} & \multicolumn{2}{c|}{65} & \multicolumn{2}{c|}{167} & \multicolumn{2}{c|}{181} & \multicolumn{2}{c|}{118} & \multicolumn{2}{c|}{45}  & \multicolumn{2}{c}{26} \\
             & 11.25 & 0.05                                & 10.70 & 0.13            & 11.06 & 0.08             & 11.33 & 0.08             & 11.47 & 0.13             & 11.52 & 0.16             & 11.82 & 0.27 \\
             & \multicolumn{2}{l|}{$N_{\rm UFG}=235$}      & \multicolumn{2}{c|}{8}  & \multicolumn{2}{c|}{12}  & \multicolumn{2}{c|}{56}  & \multicolumn{2}{c|}{63}  & \multicolumn{2}{c|}{87}  & \multicolumn{2}{c}{9}  \\
             & 11.96 & 0.06                                & 11.75 & 0.37            & 11.29 & 0.24             & 11.65 & 0.15             & 12.02 & 0.12             & 12.23 & 0.13             & 12.00 & 0.41 \\ \hline
$V_{\rm LG}$ & \multicolumn{2}{c|}{$N_{\rm RFGC-UFG}=722$} & \multicolumn{2}{c|}{67} & \multicolumn{2}{c|}{170} & \multicolumn{2}{c|}{201} & \multicolumn{2}{c|}{143} & \multicolumn{2}{c|}{67}  & \multicolumn{2}{c}{74} \\
             & 4708 & 92                                   & 5591 & 278              & 5832 & 181               & 5285 & 168               & 3920 & 166               & 3365 & 240               & 2498 & 207  \\
             & \multicolumn{2}{l|}{$N_{\rm UFG}=333$}      & \multicolumn{2}{c|}{8}  & \multicolumn{2}{c|}{16}  & \multicolumn{2}{c|}{70}  & \multicolumn{2}{c|}{80}  & \multicolumn{2}{c|}{134} & \multicolumn{2}{c}{25} \\
             & 5057 & 127                                  & 5128 & 758              & 7071 & 397               & 5467 & 243               & 5730 & 274               & 4644 & 179               & 2660 & 400  \\ \hline
$V_m$        & \multicolumn{2}{c|}{$N_{\rm RFGC-UFG}=722$} & \multicolumn{2}{c|}{67} & \multicolumn{2}{c|}{170} & \multicolumn{2}{c|}{201} & \multicolumn{2}{c|}{143} & \multicolumn{2}{c|}{67}  & \multicolumn{2}{c}{74} \\
             & 145 & 2                                     & 184 & 6                 & 178 & 4                  & 154 & 4                  & 124 & 3                  & 108 & 4                  & 81 & 5      \\
             & \multicolumn{2}{l|}{$N_{\rm UFG}=333$}      & \multicolumn{2}{c|}{8}  & \multicolumn{2}{c|}{16}  & \multicolumn{2}{c|}{70}  & \multicolumn{2}{c|}{80}  & \multicolumn{2}{c|}{134} & \multicolumn{2}{c}{25} \\
             & 131 & 3                                     & 154 & 24                & 185 & 12                 & 150 & 6                  & 139 & 6                  & 118 & 3                  & 84 & 6      \\ \hline
$g-r$     & \multicolumn{2}{c|}{$N_{\rm RFGC-UFG}=306$} & \multicolumn{2}{c|}{19} & \multicolumn{2}{c|}{84}  & \multicolumn{2}{c|}{97}  & \multicolumn{2}{c|}{51}  & \multicolumn{2}{c|}{23}  & \multicolumn{2}{c}{32} \\
             & 0.72 & 0.03                                 & 0.83 & 0.03             & 0.77 & 0.02              & 0.70 & 0.02              & 0.60 & 0.02              & 0.51 & 0.07              & 0.41 & 0.10 \\
             & \multicolumn{2}{l|}{$N_{\rm UFG}=149$}      & \multicolumn{2}{c|}{0}  & \multicolumn{2}{c|}{3}   & \multicolumn{2}{c|}{30}  & \multicolumn{2}{c|}{40}  & \multicolumn{2}{c|}{69}  & \multicolumn{2}{c}{7}  \\
             & 0.64 & 0.02                                 & \multicolumn{2}{c|}{--} & 0.95 & 0.01              & 0.80 & 0.02              & 0.60 & 0.04              & 0.57 & 0.03              & 0.42 & 0.16 \\ \hline
$W2$--$W4$   & \multicolumn{2}{c|}{$N_{\rm RFGC-UFG}=687$} & \multicolumn{2}{c|}{67} & \multicolumn{2}{c|}{169} & \multicolumn{2}{c|}{196} & \multicolumn{2}{c|}{135} & \multicolumn{2}{c|}{59}  & \multicolumn{2}{c}{61} \\
             & 5.11 & 0.03                                 & 4.42 & 0.12             & 4.81 & 0.06              & 5.16 & 0.05              & 5.34 & 0.05              & 5.43 & 0.07              & 5.71 & 0.08 \\
             & \multicolumn{2}{l|}{$N_{\rm UFG}=301$}      & \multicolumn{2}{c|}{7}  & \multicolumn{2}{c|}{15}  & \multicolumn{2}{c|}{64}  & \multicolumn{2}{c|}{78}  & \multicolumn{2}{c|}{118} & \multicolumn{2}{c}{19} \\
             & 5.16 & 0.04                                 & 5.03 & 0.61             & 4.77 & 0.18              & 4.93 & 0.09              & 5.17 & 0.06              & 5.26 & 0.05              & 5.64 & 0.13 \\ \hline
\end{tabular}
\end{table*}

Table~1 lists the mean values and errors in mean for the observed
characteristics of UFG  and \mbox{RFGC--UFG} galaxies depending on
the type. The bottom line corresponds to ultra-flat galaxies, while
the top line lists    flat galaxies. The few cases of $T=2$ and $T=9$
are respectively combined with the     $T=3$ and $T=8$ types.
 From a
comparison of the data presented      the following
conclusions can be made.

\begin{list}{}{
\setlength\leftmargin{2mm} \setlength\topsep{2mm}
\setlength\parsep{0mm} \setlength\itemsep{2mm} }
  \item [$\bullet$] The mean apparent axial ratio shows the expected increase from the early-type to late-type spirals. The peak of the distribution for UFG galaxies both in the blue and in the red bands falls on   $T=7$, whereas for
RFGC--UFG galaxies the  maximum is fixed at  $T=6$.

 \item[$\bullet$]
Among all the morphological types, ultra-flat galaxies look somewhat
fainter than simply flat galaxies (RFGC--UFG). The mean difference in
the apparent magnitudes amounts to \mbox {$+0\fm52\pm0\fm05$} in the
$B$-band and  $+0\fm71\pm0\fm08$ in the $K_s$-band. The reason for
this difference could be  internal extinction in strongly inclined
disks. However, then the difference in the infrared band would be
significantly smaller than in the  blue band. A more likely cause is
hidden selection effects due to methodological features of the
photometry of extended low-contrast galaxy images.

 \item[$\bullet$]  The average
radial velocity (i.e., distance) of ultra-flat galaxies does not
differ much  from the average for the RFGC--UFG-objects. Also, a
small difference is found for the mean rotation amplitude $V_m$,
which
 is closely correlated with the luminosity of the galaxy. In other words, both
samples refer  to  about the same volume of space.

 \item[$\bullet$]  A significant portion of studied objects lie in the zone of the
SDSS survey (DR13), what allows   to compare their average optical
color indices $\langle g-r \rangle$. According to the  table data,
the color differences of UFG and RFGC--UFG galaxies are small, which
does not impose any restriction on the nature of internal extinction
in thin disks.

 \item[$\bullet$]  More than 90\% of
RFGC galaxies are detected in four  infrared bands of the WISE sky
survey~\cite{jar2012:Melnyk_n_en}: $W1$, $W2$, $W3$, $W4$
($3.38~\mu{\rm m}$, $4.6~\mu{\rm m}$, $12.33~\mu{\rm m}$ and
$22.00~\mu{\rm m}$). Here the most long-wavelength band is the most
sensitive to the thermal flux of dust, re-emitting  the total stellar
flux. The   $W4$ value is an additional indicator of the amount of
dust and  integral SFR in a galaxy. As can be seen from the average
values of $\langle W2-W4 \rangle$, the color index increases steadily
from the early to late  morphological types.
 Between the values of $\langle W2-W4 \rangle$ in general, no significant differences were  observed in the UFG and RFGC--UFG galaxies.
However, within each morphological type ultra-flat
galaxies look somewhat more ``blue''
 (less
dusty) than the others.
\end{list}

\section{ACCOUNTING FOR INTERNAL EXTINCTION IN SPIRAL GALAXIES}

The presence of dust in the disk of a spiral galaxy reduces its
integral luminosity. The internal extinction effect is the most
strongly manifested in edge-on galaxies,   and grows from the
infrared to ultraviolet spectral regions. The   distribution of
bright blue stars and dust clouds in the disk of a spiral galaxy is
extremely uneven and can not be described by a simple model of
plane-parallel layers. Therefore, so far there is no reliable and
universal scheme of accounting for internal extinction.

The formulation of the corrections to the apparent magnitude of a
galaxy for its inclination, adopted in various editions of The
Reference Catalogue of Galaxies (see, e.g.,~\cite{vau1991:Melnyk_n_en}) and in
the HyperLEDA, changed over time. Its main drawback was that it
ignored the dependence of the  extinction magnitude on the luminosity
of a spiral galaxy, to what   Tully et al.~\cite{tul1998:Melnyk_n_en} and
Verheijen  and  Sancisi~\cite{ver2001:Melnyk_n_en} have paid due attention.
According to~\cite{ver2001:Melnyk_n_en},  weakening of the integral magnitude of
galaxies in the  $B$-band is expressed by:
\begin{equation}
A^i_B = [1.54 +2.54\,(\log V_m -2.2)]\,\log(a/b)
\end{equation}
for $V_m > 43$~km\,s$^{-1}$, where $a/b$ is the apparent axial ratio,
and $V_m$ is  the amplitude of rotation in km\,s$^{-1}$; $A^i_B =  0$
for \mbox{$V_m <43$~km\,s$^{-1}$}. According to~(1),  the extinction
steadily increases with increasing $V_m$  or luminosity. Basically
this formula reasonably describes   internal extinction in spiral
galaxies, giving on the average a lower correction for the extinction
than the RC3 and LEDA.

In their recent publication~\cite{dev2016:Melnyk_n_en} Devour and Bell examined
the integral features of 78\,720 SDSS galaxies with redshifts $z<0.1$
and apparent red magnitudes $m_r<17\fm7$. Comparing different ways of
accounting for internal extinction, the authors came to the
conclusion that the  schemes of corrections for the inclination,
proposed in~\cite{tul1998:Melnyk_n_en} and~\cite{mal2009:Melnyk_n_en} overestimate the
extinction for the brightest galaxies. According to~\cite{dev2016:Melnyk_n_en},
internal extinction increases with increasing luminosity of a spiral
galaxy not monotonely, but has a peak at the maximum absolute
magnitude \mbox{$M_K=-21.7$}. Based on the data of Fig.~12
from~\cite{dev2016:Melnyk_n_en},  we found that the value of internal extinction
in the   $B$-band
 for a spiral galaxy with the absolute magnitude   $M_K$
at  $H_0=73$~km\,s$^{-1}$\,Mpc$^{-1}$ can be represented by a
parabolic relation
\begin{equation}
A_B^i = [0.80 - 0.0584 ( M_K^{\rm corr} + 21.7)^2]\,\log (a/b)_r,
\end{equation}
where $(a/b)_r$ is the apparent axial ratio of the galaxy measured in
the RFGC catalog at  the red images of the   Palomar  Observatory Sky
Atlas. Expression~(2) is applicable to the range of  \mbox{$-17.7 >
M_K>-25.7$}, and outside of it the negative values of $A^i$ are
replaced by zero. For example, for the SMC dwarf galaxy with  $M_K=
-18.8$, $V_m = 46$~km\,s$^{-1}$ and \mbox{$(a/b)=1.56$},
relationships~(1) and~(2)  give similar values: \mbox{$A^i = 0\fm03$
and $0\fm06$} respectively. However, for an ultra-flat galaxy with
$(a/b)=10$,\linebreak {$V_m = 200$~km\,s$^{-1}$} and $M_ K= -24.2$
the difference of the corrections for  $A^i$   becomes significant:
$1\fm80$  and $0\fm44$. Here for the transition from  $V_m$ to  $M_K$
we used the calibration ratio from~\cite{tul1998:Melnyk_n_en}
\begin{equation}
M_K^{\rm corr}= -23.29 - 8.78\,(\log V_m - 2.20).
\end{equation}
At that, the range of positive values of the  $A^i$  correction
corresponds to the $V_m$  interval from  39~km\,s$^{-1}$ to
300~km\,s$^{-1}$.

The amplitude of  rotation $V_m$, which we adopted from the HyperLEDA, is now
known for 85\% of the \mbox{RFGC galaxies}. In the absence of the data on
$V_m$ we calculated this value from the empirical regression between
$V_m$~(km\,s$^{-1}$)  and the morphological type of a given galaxy:
\begin{equation}
V_m = -20.36\, T + 254.
\end{equation}

We made the recalculation of the correction for the internal  $A_B^i$
and external (Galactic) extinction $A^G$ from the $B$-band to the
FUV, $K_s$  and  $W1$-bands using the relations:
\begin{eqnarray}
A^t_{\rm FUV}& = & 1.930\,(A_B^i + A^G), \nonumber \\
A^t_{K_s}& = & 0.083\,(A_B^i + A^G),\\
A^t_{W1}& = & 0.052\,(A_B^i+A^G), \nonumber
\end{eqnarray}
where the parabolic dependence~(2)  was used to take into account the
internal extinction.

\section{FUV FLUXES AND SFR}

We expressed the integral SFR of a galaxy through its apparent
magnitude in the FUV-band, corrected for the extinction as
in~\cite{mel2015:Melnyk_n_en}:
\begin{equation}
\log\,SFR = 2.78 - 0.4\, m_{c,\,{\rm FUV}} + 2\,\log D,
\end{equation}
where  the distance $D = V_{\rm LG} /73$ is expressed in~Mpc, and
$SFR$---in $M_{\odot}\,{\rm yr}^{-1}$. According
to~\cite{ken1998:Melnyk_n_en,lee2009:Melnyk_n_en}, this relation
fixes the SFR at the characteristic timescale of about   $10^8$~yrs
determined by the radiation of young blue stars.

\begin{figure}[]
\setcaptionmargin{5mm} \onelinecaptionstrue \captionstyle{normal}
\includegraphics[angle=0, width=0.9\columnwidth]{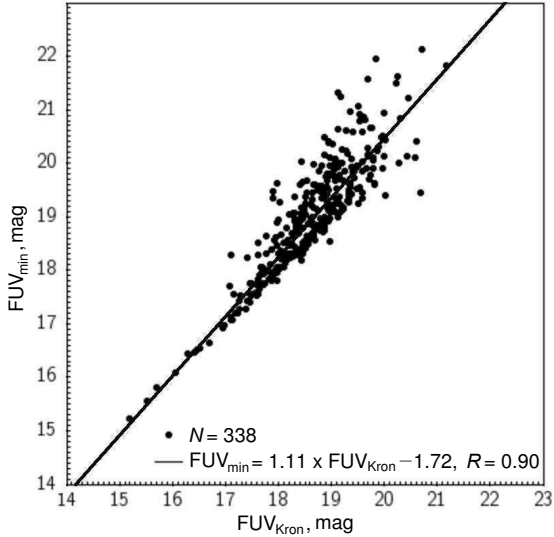}
\caption{A comparison of the of the FUV photometry data in the elliptical
Kron and circular apertures. The figure shows the
regression parameters.} \label{fig3:Melnyk_n_en}
\end{figure}

The sky survey made with the GALEX Space
Telescope~\cite{mar2005:Melnyk_n_en}, registered the
FUV\mbox{-}fluxes for approximately 75\% of  galaxies in our sample.
In the softer  NUV-range the percentage is slightly higher (84\%),
but we will further restrict to using only the FUV\mbox{-}magnitudes,
as we did in our previous
papers~\cite{kar2013a:Melnyk_n_en,kar2013b:Melnyk_n_en,mel2015:Melnyk_n_en}.
The images of thin spiral disks in the FUV and NUV-bands look
low-contrast and shredded, often breaking into a chain of separate
condensations with individual $m_{\rm FUV}$ and $m_{\rm NUV}$
estimates. Determining the SFR in a galaxy we focused on the values
of  Kron elliptical aperture magnitudes, which are available for
almost half of the objects of our sample. For the rest, we used the
so-called $m_{\rm min}$, corresponding to the FUV-flux from the
brightest concentration in the galaxy. This technique allowed to
avoid the manual analysis of numerous individual images having a low
angular resolution (about $5\arcsec$)  and low contrast. The
distribution of our galaxies by the FUV-magnitudes $m_{\rm Kron}$ and
$m_{\rm min}$ is shown in Fig.~\ref{fig3:Melnyk_n_en}. The relation
between them is described by a regression
\begin{equation}
m_{\rm FUV,\,Kron} = 0.9\,m_{\rm FUV,\,min} +1.55
\end{equation}
with the correlation coefficient  $R=0.90$. On the average the transition from $m_{\rm min}$ to $m_{\rm Kron}$ required a correction by about $-0\fm4$ at the
typical dispersion of around $0\fm5$.
Further in the paper we use
Kron's FUV-magnitudes directly measured in the elliptic
apertures or recomputed  from $m_{\rm min}$  by relation~(7).
Thus, the integral SFR was determined by
 relation~(6) accounting for the corrections~(2) and ~(5).

It is known that the integral SFR of a galaxy quite closely
correlates with its total stellar mass, forming the so-called ``main
sequence'' on the \linebreak $\log SFR \propto \log M_*$ diagram.
 Therefore,
further for each galaxy  in our sample, we determined its stellar mass
using   different approaches.

\section{GALAXY STELLAR MASS ESTIMATION}

Integral  stellar mass of a galaxy $M_*$ is usually determined by the
spectral energy distribution (SED), specifying a particular shape of
the initial    stellar mass function (IMF) by
Salpeter~\cite{sal1955:Melnyk_n_en},
Kroupa~\cite{kro2016:Melnyk_n_en} or
Chabrier~\cite{cha2003:Melnyk_n_en}. In practice,  $M_*$ is more
frequently estimated using  integral luminosity of a galaxy in
certain band, taking a fixed  mass-to-luminosity ratio. We calculated
the stellar mass from  the galaxy  $K$-band luminosity  at the value
of \mbox{$M_*/L_K
=1(M_{\odot}/L_{\odot})$}~\cite{bel2003:Melnyk_n_en} and  \mbox
{$M_{\odot,\,K} = 3\fm28$}~\cite{bin1998:Melnyk_n_en}. As noted
in~\cite{mcg2015:Melnyk_n_en}, the  $M_*/L_K$  relation has not yet
been fixed very reliably and most likely lies in the range of ${\mbox
[0.5-1.0]} M_{\odot}/L_{\odot}$.

The apparent $K$-magnitude of the galaxy can be determined in several ways.
\begin{list}{}{
\setlength\leftmargin{2mm} \setlength\topsep{2mm}
\setlength\parsep{0mm} \setlength\itemsep{2mm} }
  \item 1)  The infrared survey the entire sky 2MASS~\cite{scr2006:Melnyk_n_en} contains
automatic evaluations of $K_s$-magnitudes for 73\%  of  galaxies in
our sample. However, due to short exposures (approximately 8~s) the
survey is insensitive to bluish structures of low surface brightness.
As a result of underestimation of the real size of thin disks and
their frequent  separation into a few fragments, the integral
$K_s$\mbox{-}fluxes of many RFGC galaxies are systematically
underestimated, particularly for faint galaxies with $K_s > 12^{\rm
m}$.

 \item 2)  The  WISE infrared sky survey~\cite{jar2012:Melnyk_n_en} has the magnitude estimates in four infrared bands: $W1$, $W2$, $W3$  and  $W4$ for 94\% galaxies
under consideration. This survey, as well as 2MASS suffers from the
flux underestimation from the periphery of diffuse galaxies. The
apparent magnitude of RFGC galaxies in the $W1$ band, closest to
$K_s$, shows a tight correlation with    $K_s$; the ratio between
them has the form
\begin{equation}
K_s=0.99\,K_{W1}^c - 0.06,
\end{equation}
where  $K_{W1}^c = m_{W1} -0.83$, the correlation coefficient
$R=0.90$, the number of galaxies  \mbox{$N=813$}.  This
\mbox{$K_s$-magnitude}, derived from the $W1$-magnitude  shall be
further referred to as  $K_W$.
 \item 3) According to~\cite{jar2003:Melnyk_n_en}, the mean color index of the galaxy depends
on its morphological type as
 \begin{equation}
 \langle B-K \rangle_{\rm corr} = 4.60 - 0.25\,T
 \end{equation}
at  $T = 2, 3,... 9$. Using this relation and taking into account the
corrections for the extinction~(2) and (5), we determined the
$K$-magnitudes for all the RFGC galaxies. These values will be
further denoted as $K_B$. The regression line for them and the \mbox
{$K_s$-magnitude}, constructed from 837 galaxies, has the form
\begin{equation}
K_s=0.56\,K_B+4.87.
\end{equation}
 \item 4) We can also estimate the absolute \mbox{$K$-magnitude} from the calibration
 relation~(3) considering the additional relation~(4) for the galaxies
without direct   $V_m$ measurements.
\end{list}

\begin{figure}[]
\setcaptionmargin{5mm} \onelinecaptionstrue \captionstyle{normal}
\includegraphics[angle=0, width=0.8\columnwidth]{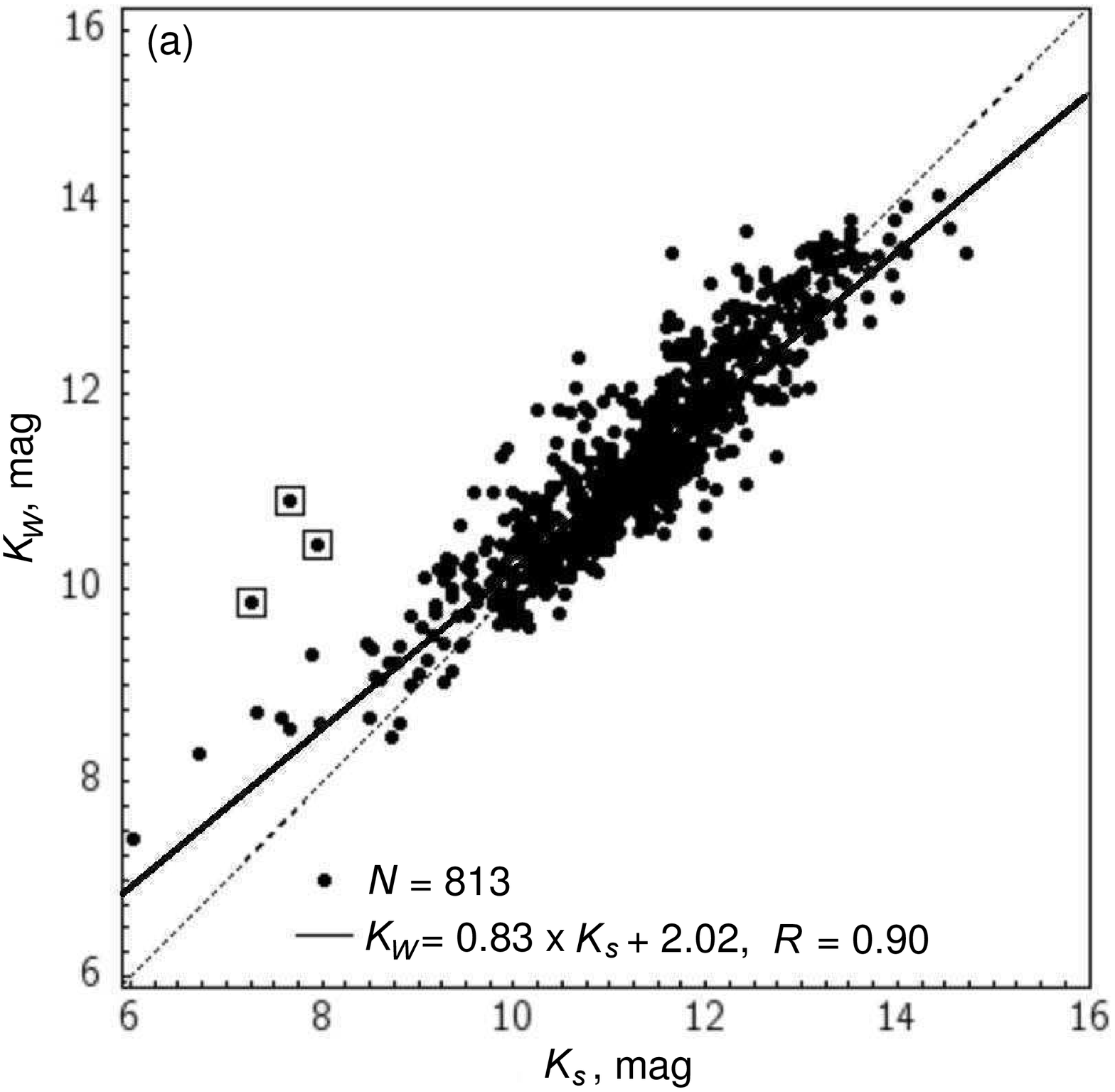}
\includegraphics[angle=0, width=0.8\columnwidth]{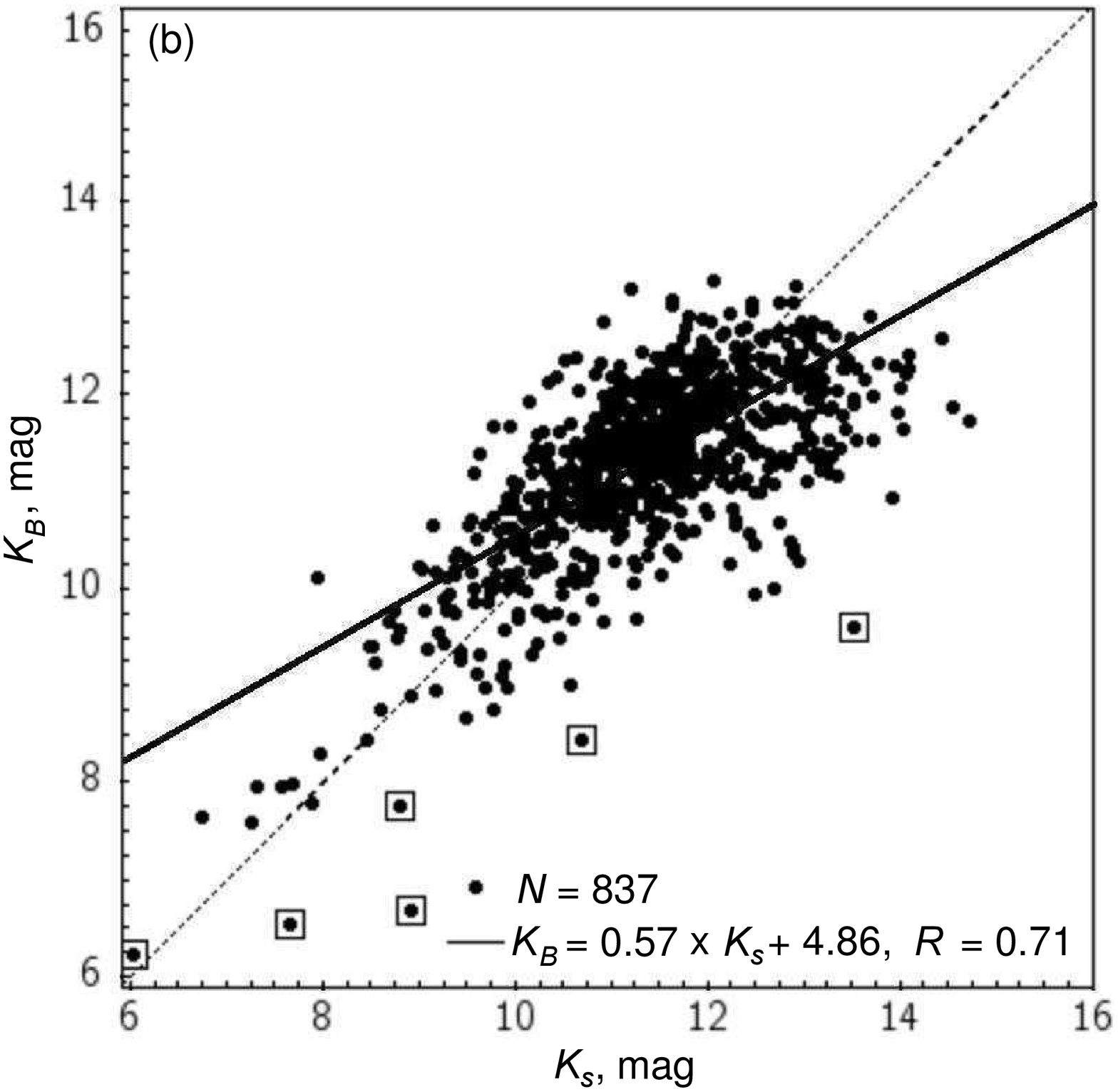}
\includegraphics[angle=0, width=0.8\columnwidth]{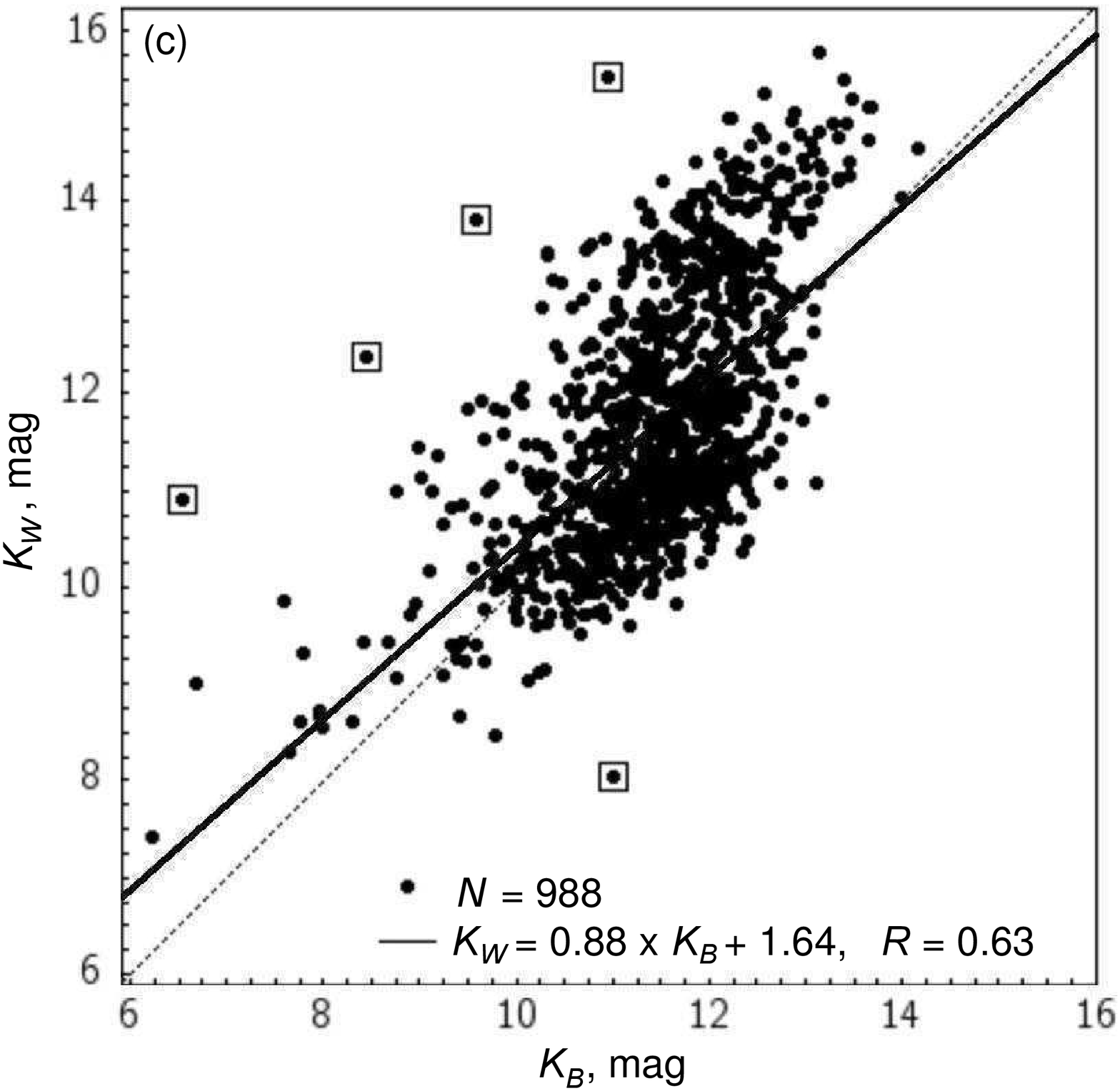}
\caption{Mutual dependences of the calculated and observed magnitudes
for flat and ultra-flat galaxies (from top to bottom): $K_W$~vs~$K_s;
K_B$~vs~$K_s; K_W$~vs~$K_B$. The figures show the corresponding
regression parameters and the number of galaxies. The squares
outline the objects strongly diverging from the regression line.}
\label{fig4:Melnyk_n_en}
\end{figure}

Two-dimensional distributions of galaxies from our sample based on
\{$K_s, K_W$\}, \{$K_s, K_B$\} and \linebreak \{$K_W, K_B$\} are
shown in three panels of Fig.~\ref{fig4:Melnyk_n_en}. The solid and dashed lines
denote the regression line and the diagonal, respectively. Apart from
the
 main array of galaxies, each panel contains about 1\% of
cases with large deviations from the regression line.
In Fig.~\ref{fig4:Melnyk_n_en} such objects are outlined by squares. Additional
analysis shows that the main reasons for these deviations
are the infrared photometry errors, when there is a loss of flux
the periphery of a given galaxy (the cases of RFGC\,566, 1295, 2245, 2315, 2335,
3854), or a bright star is projected on the galaxy   (RFGC\,668,
1049, 1340, 2140, 3357). In our experience, the most reliable measurement
of stellar mass in thin spiral edge-on galaxies
 is to use the   \mbox{$K_B$-magnitude}.

\section{SPECIFIC AND EFFECTIVE SFR}

\begin{figure*}[]
\setcaptionmargin{5mm} \onelinecaptionstrue \captionstyle{normal}
\includegraphics[angle=0, width=0.9\columnwidth]{Melnyk_fig5a.eps}
\includegraphics[angle=0, width=0.9\columnwidth]{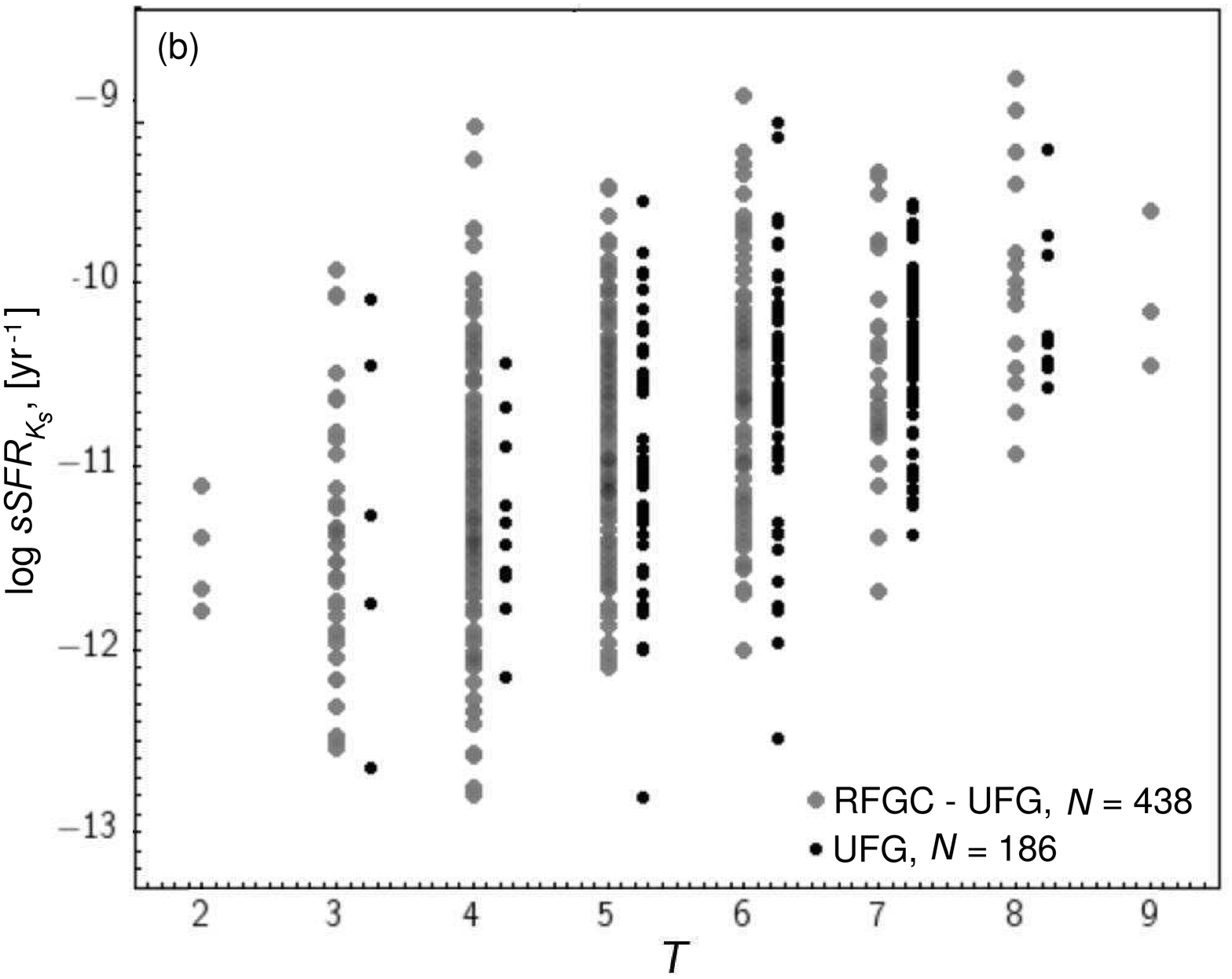}
\includegraphics[angle=0, width=0.9\columnwidth]{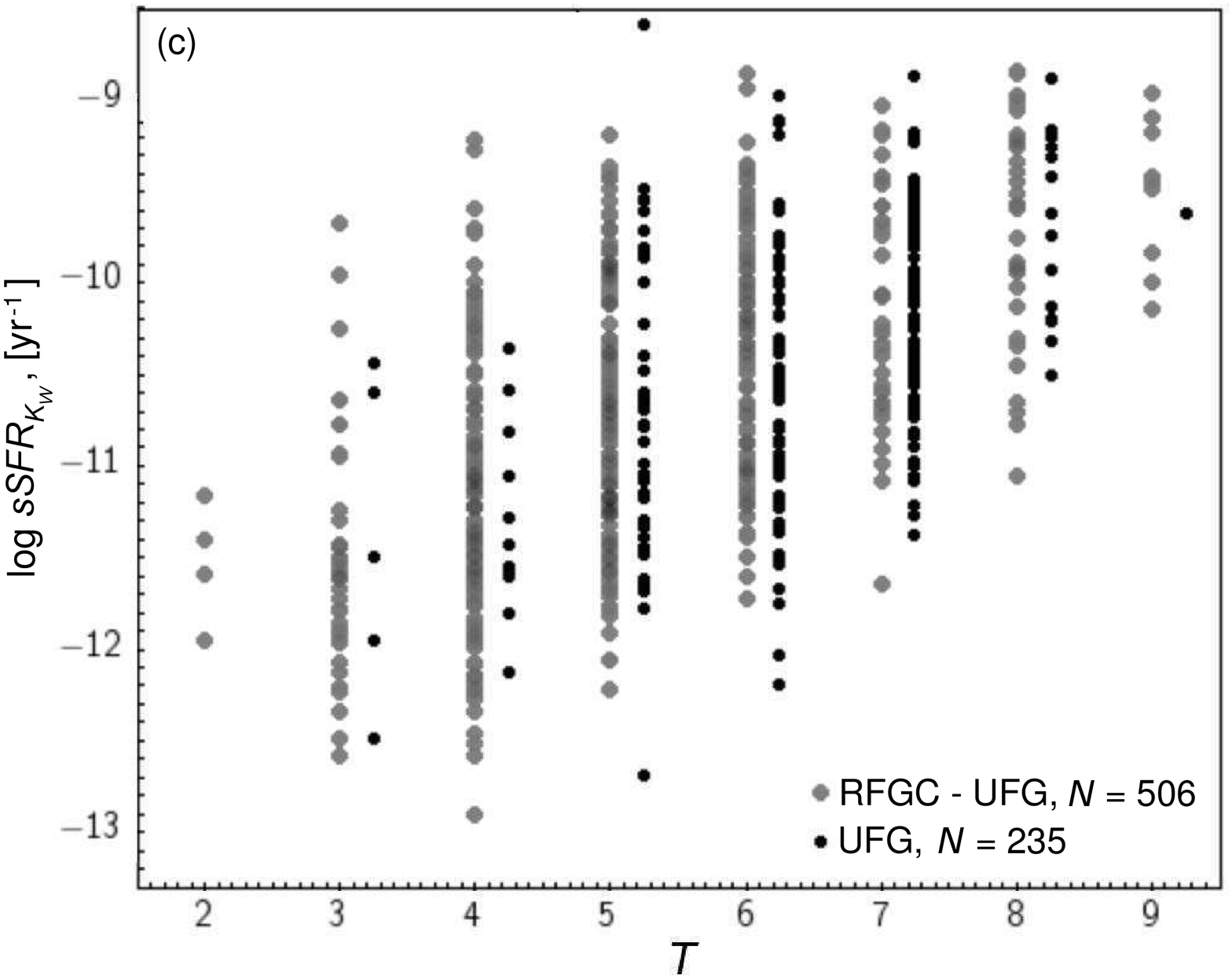}
\includegraphics[angle=0, width=0.9\columnwidth]{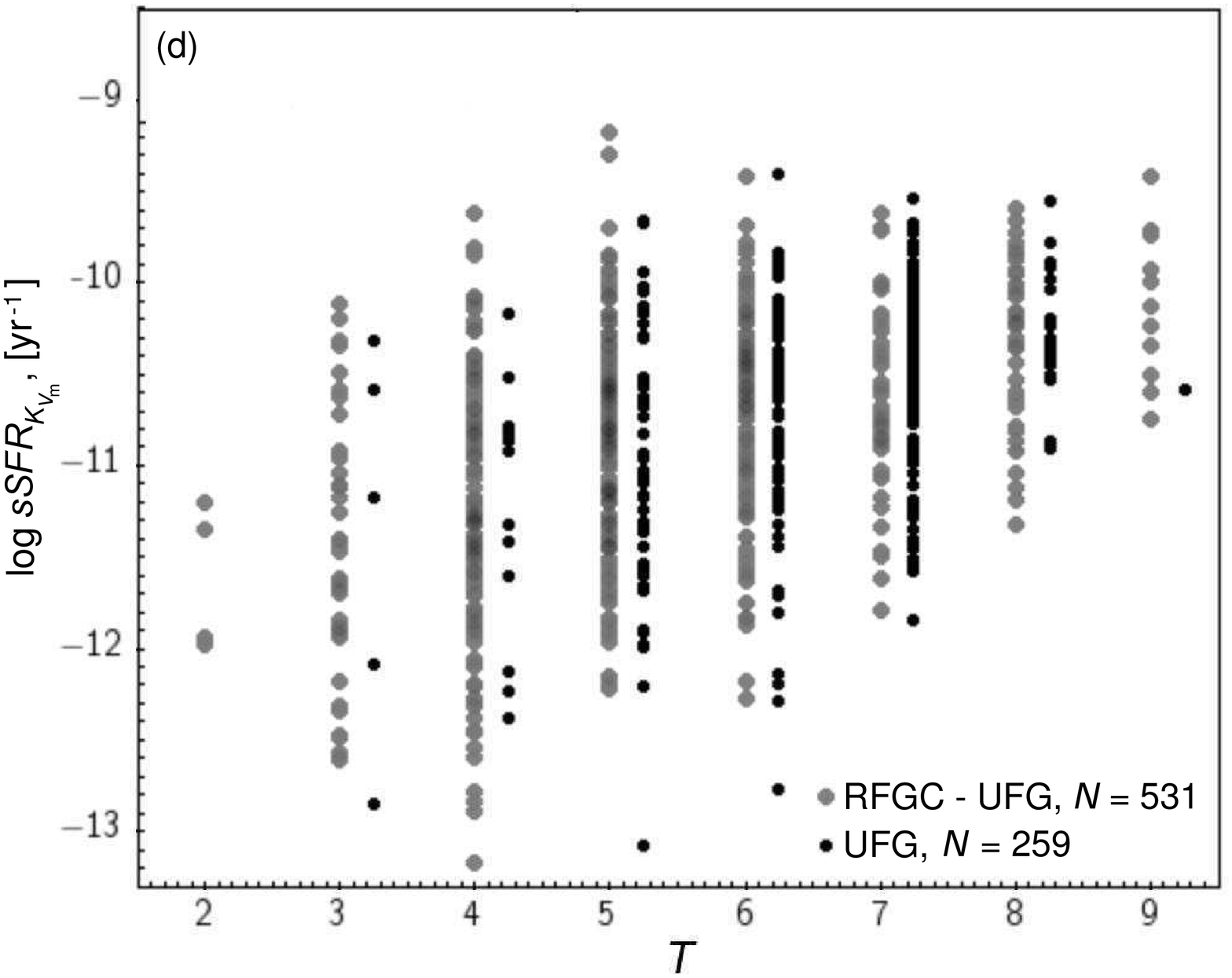}
\caption{The dependence of specific star formation rate
$\log\,sSFR$  on the morphological type of the galaxy $T$  for \mbox
{RFGC--UFG} and  UFG galaxies---gray and dark circles
respectively. Stellar mass of the galaxy was determined from its
$K$-luminosity at $M_*=1 M_{\odot}/L_{\odot}$. The $K$-magnitudes of
galaxies were evaluated in four different ways:  $K_B$---from the
$B$-magnitude  and morphological type, $K_s$---according to the
2MASS-survey, $K_w$---from the $W1$-magnitude of the WISE-survey,
$K_{V_m}$---from relation~(3) between the rotation amplitude $V_m$ and
absolute $K$-magnitude.} \label{fig5:Melnyk_n_en}
\end{figure*}

Dependences of the specific star formation rate  $sSFR = SFR/M_*$  on
the morphological type of the galaxy $T$ for the UFG  (dark circles)
and RFGC--UFG (gray circles) disks are presented in four panels of
Fig.~\ref{fig5:Melnyk_n_en}. Stellar mass was identified with the
$K$-luminosity \mbox{$(M_* = L_K)$}, and \mbox{$K$-magnitudes} were
determined in four variants described above. For a better
visualization, the data for the UFG sample are shifted to the right
on the horizontal axis. The corresponding number of galaxies is
indicated in the corner of each panel.

As might be expected, specific SFR on the average
increases from early to late spirals. However, comparing the diagrams
we can see that the most clear dependence $\log\,sSFR$ on the
type manifests itself when  $K_B$-magnitude is used to determine the stellar mass. No specific differences in the $sSFR$  between  the UFG  and
\mbox{RFGC--UFG} galaxies are  observed.

 \begin{figure*}[]
 \setcaptionmargin{5mm} \onelinecaptionstrue \captionstyle{normal}
\includegraphics[angle=0, width=0.9\columnwidth]{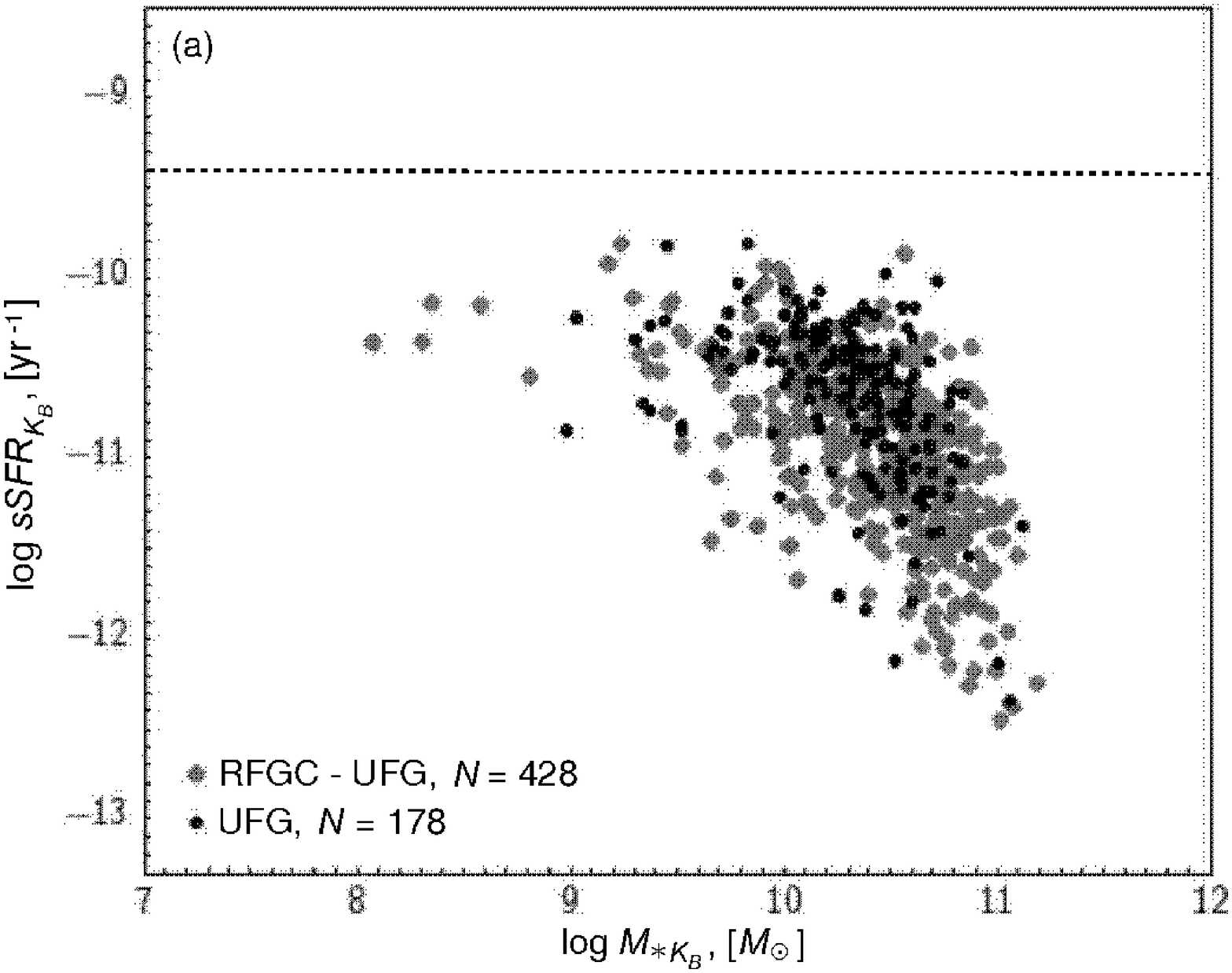}
\includegraphics[angle=0, width=0.9\columnwidth]{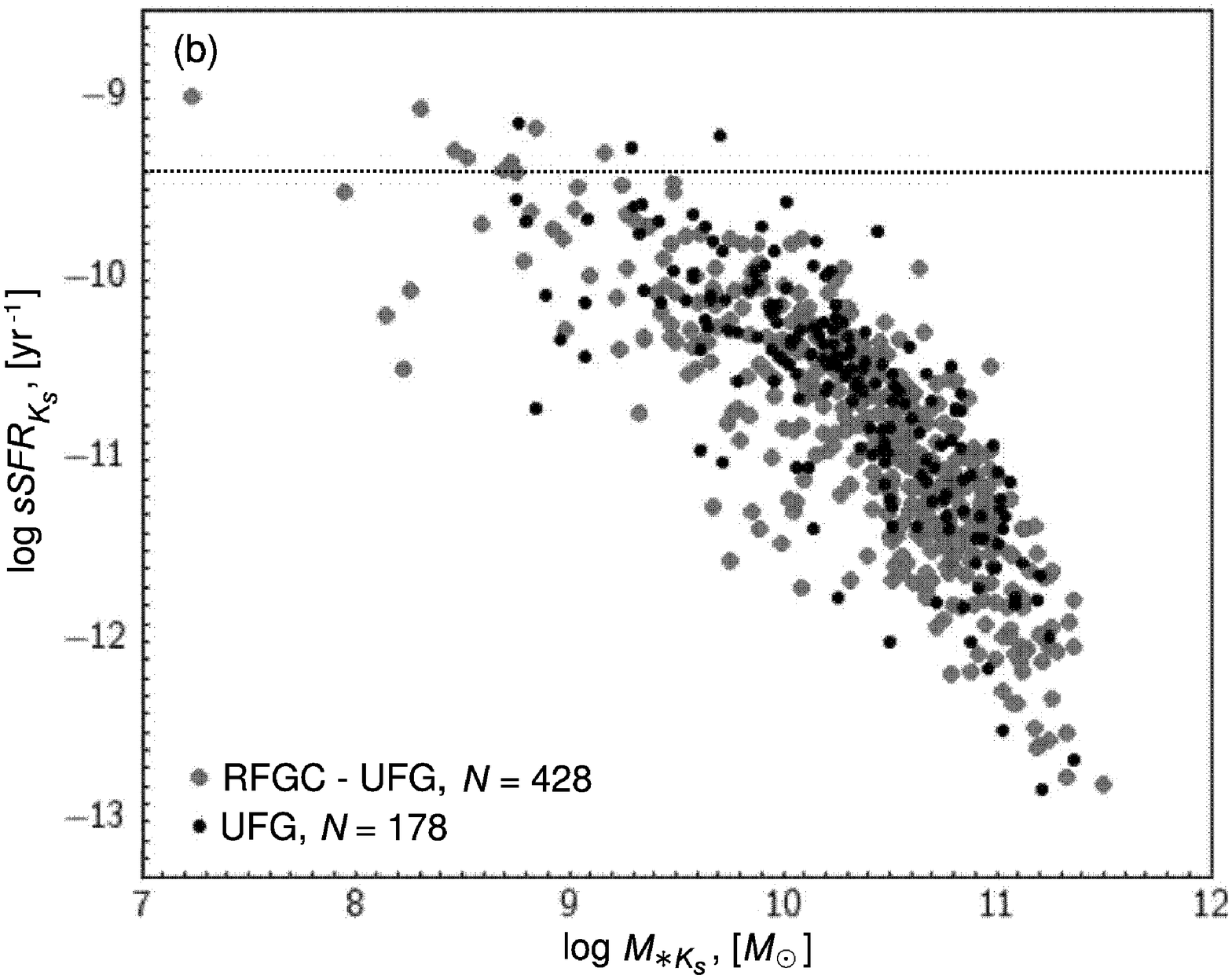}
\includegraphics[angle=0, width=0.9\columnwidth]{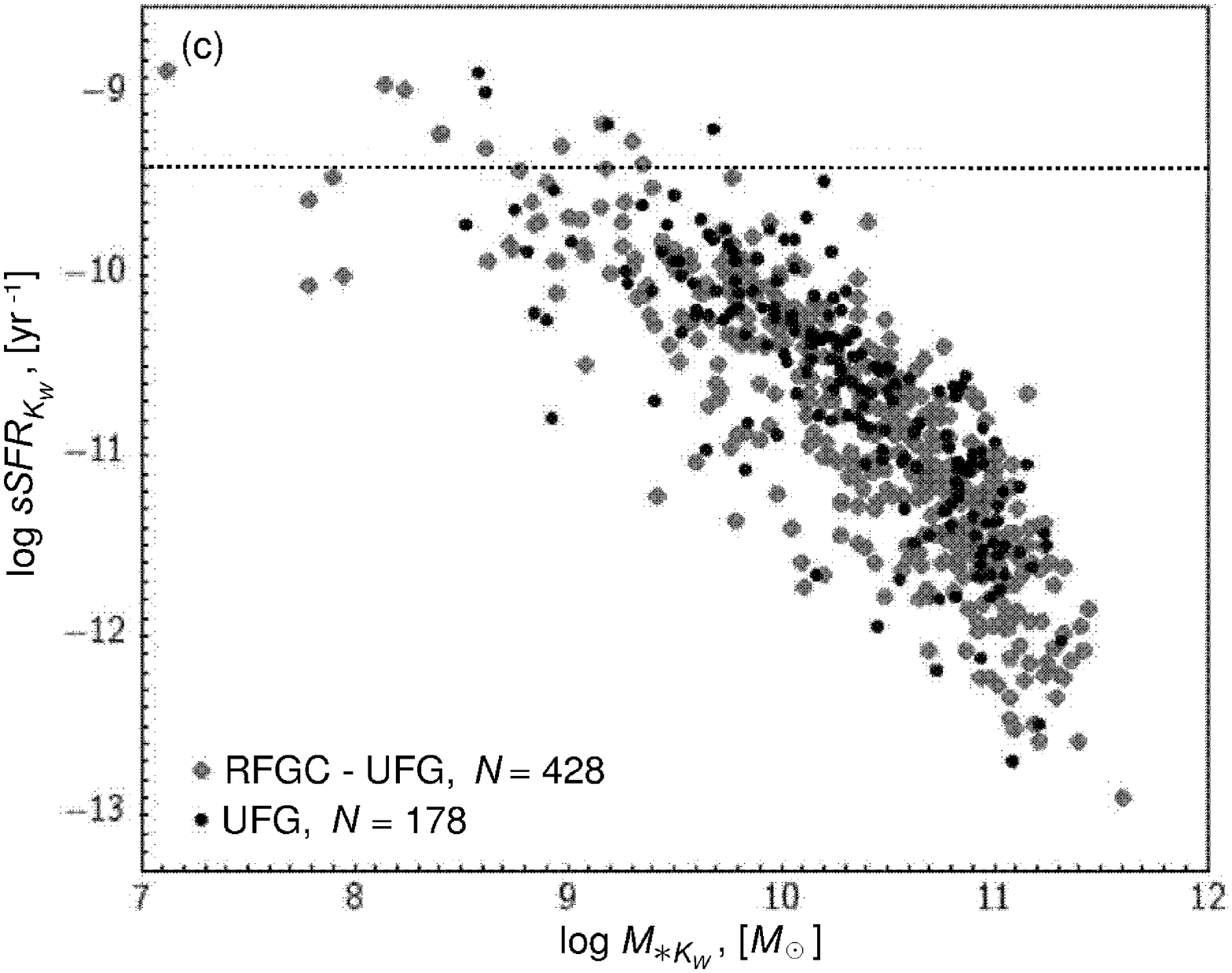}
\includegraphics[angle=0, width=0.9\columnwidth]{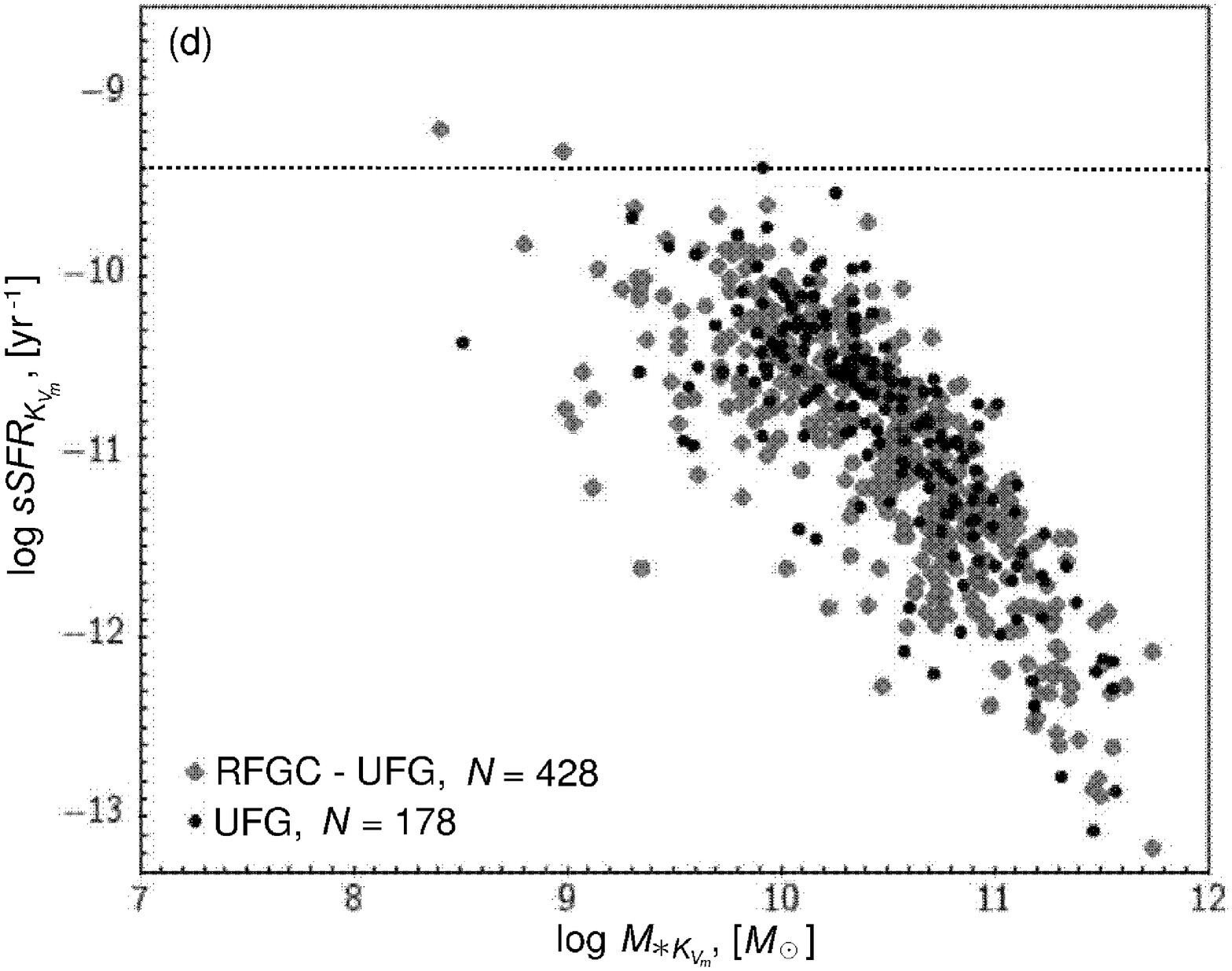}
\caption{The specific SFR, depending on the
luminosity (stellar mass) of the galaxy. Gray circles on the panels
 depict the  RFG--UFG subsample galaxies  ($N =
428$), and the dark circles---UFG galaxies ($N = 178$).
Luminosity  (stellar mass) was determined in four ways,
ÿanalogous to those in Fig.\ref{fig5:Melnyk_n_en}.} \label{fig6:Melnyk_n_en}
\end{figure*}

Figure~\ref{fig6:Melnyk_n_en}  represents the     $\log\,sSFR$ dependence on the
stellar mass for the four   $K$-luminosity estimation alternatives.
The designations of flat and ultra-flat galaxies are the same as in
the previous figure. To properly compare the data of different
panels, we presented there only 178 UFG galaxies and 428
RFGC\mbox{--}UFG galaxies, for which the {$K$-magnitude} estimates
are available by all four methods. The horizontal line in all panels
corresponds to the value of $\log\,sSFR=-9.4$ we marked
earlier~\cite{kar2013a:Melnyk_n_en,kar2013b:Melnyk_n_en,mel2015:Melnyk_n_en} as a certain
quasi-``Eddington'' limit for the star formation intensity at the
present epoch. Above this limit there are only a few galaxies, the
$K$-luminosity (stellar mass) of which is underestimated from the
infrared photometry.

As it  can be seen, the behavior of dependences for flat and ultra-flat galaxies
is approximately the same. It should be noted that the stellar mass of the
galaxy is present on both scales of  Fig.~\ref{fig6:Melnyk_n_en} panels,
therefore the inevitable  errors in the  $M_*$ due to  unreliable
photometry (sometimes reaching the order of magnitude) stretch
the observed distribution cornerwise \mbox{$\Delta \log\,sSFR =
-\Delta \log\,M_*$}. This is why the maximal $sSFR$ values
prove to be in galaxies with a minimal mass estimate.
Figure~\ref{fig6:Melnyk_n_en} once again demonstrates that the smallest scatter of
 $sSFR$ estimates is typical for the case when the $B$-band photometry
is used. Further on, from four ways of determining the
\mbox{$K$-magnitude} for RFGC galaxies we would prefer
the $B$-band photometry.

 \begin{figure}[]
\setcaptionmargin{5mm} \onelinecaptionstrue \captionstyle{normal}
\includegraphics[angle=0, width=0.9\columnwidth]{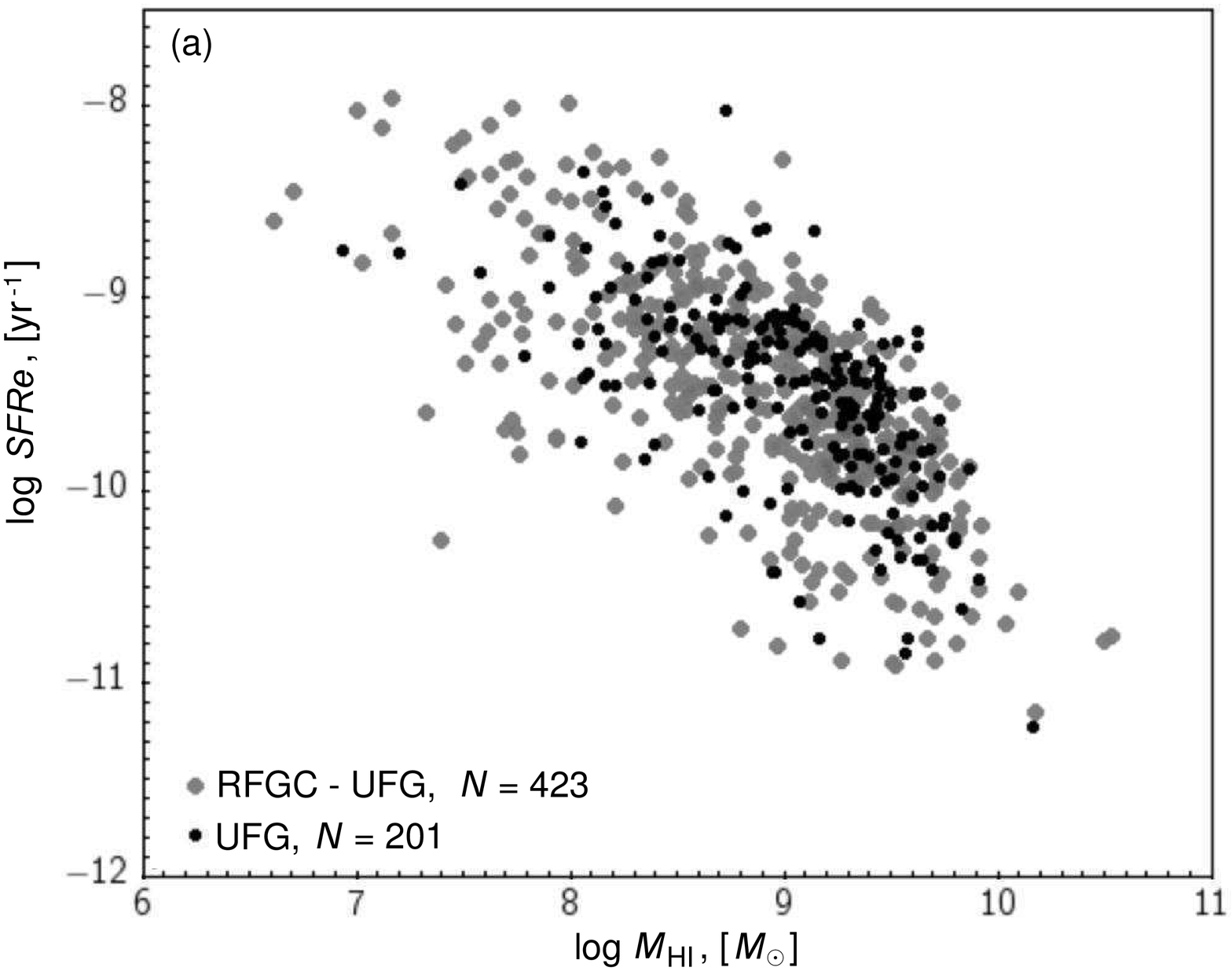}
\includegraphics[angle=0, width=0.9\columnwidth]{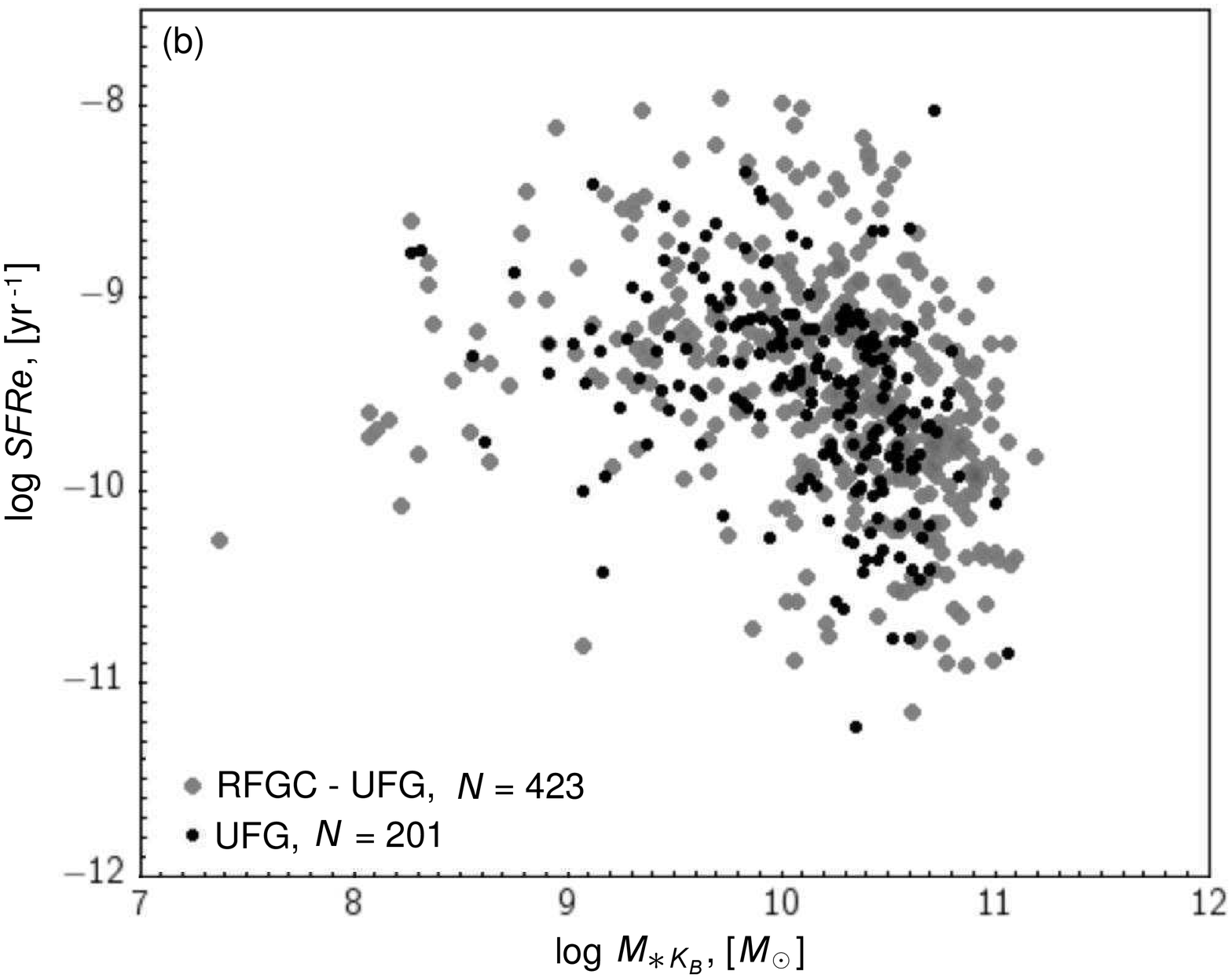}
\caption{The effective SFR as a function of (a)~hydrogen and
(b)~stellar mass of a galaxy. Stellar masses were determined from the
 apparent total magnitude $B_t$  (see Section~5).} \label{fig7:Melnyk_n_en}
\end{figure}

\begin{figure}[]
\setcaptionmargin{5mm} \onelinecaptionstrue \captionstyle{normal}
\includegraphics[angle=0, width=0.9\columnwidth]{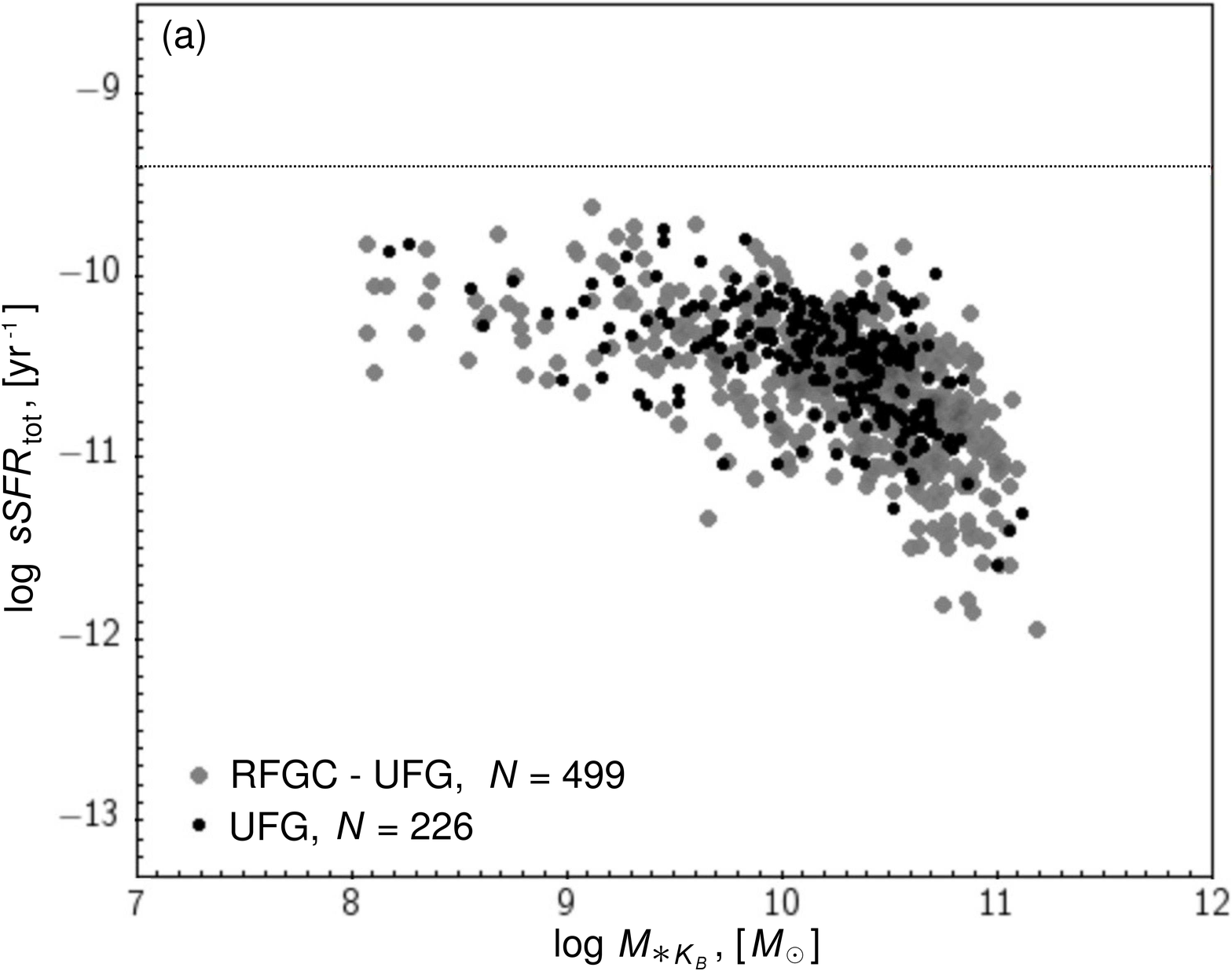}
\includegraphics[angle=0, width=0.9\columnwidth]{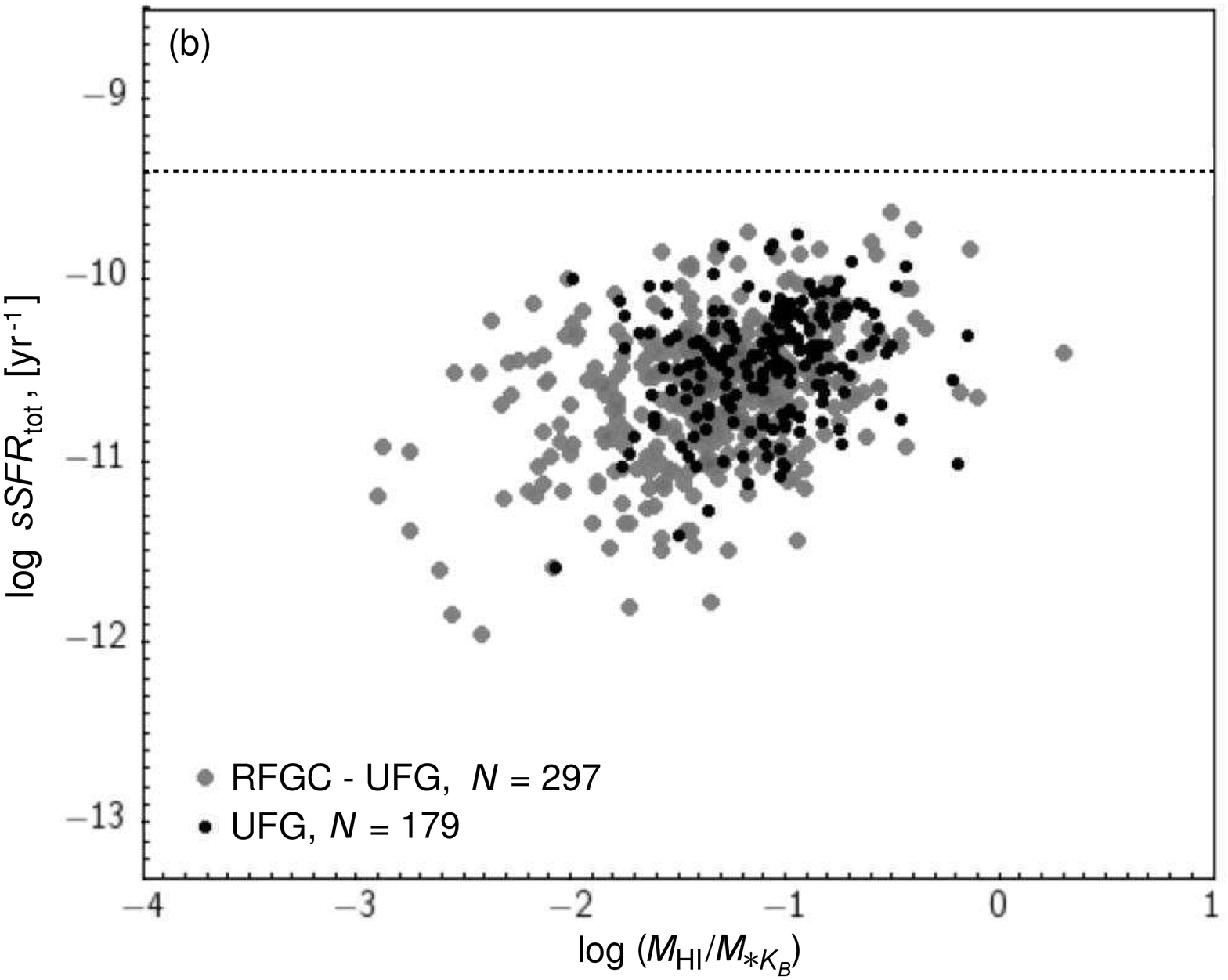}
\caption{Full (FUV\,+\,IR) relative SFR
depending on the stellar mass of the galaxy,  determined from the
 $K_s$-luminosity via  the apparent magnitude $B_t$~(a) and depending on the
value of  the hydrogen-to-stellar mass ratio~(b).}
\label{fig8:Melnyk_n_en}
\end{figure}

\begin{figure}[]
\setcaptionmargin{5mm} \onelinecaptionstrue \captionstyle{normal}
\includegraphics[angle=90, width=0.9\columnwidth]{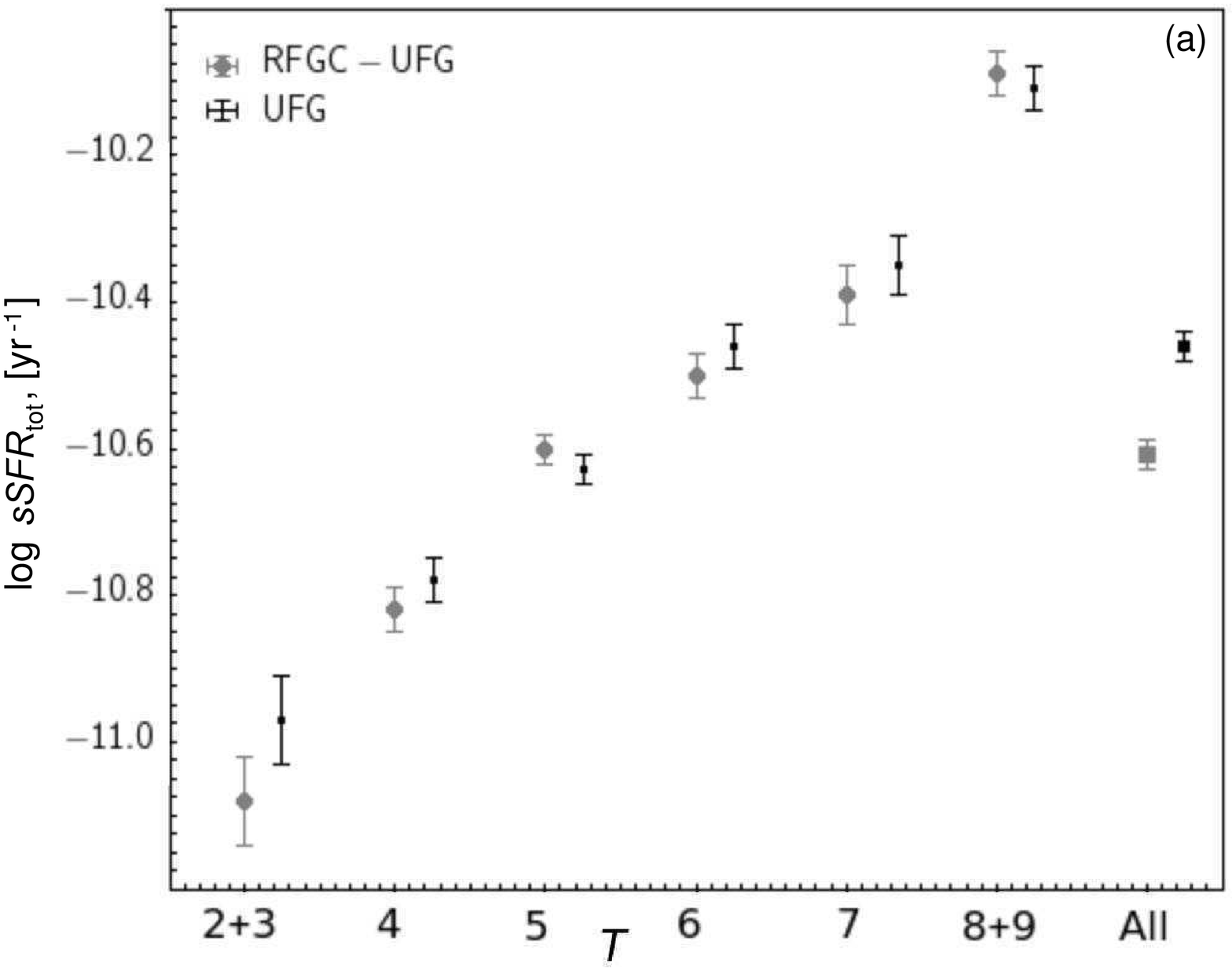}
\includegraphics[angle=90, width=0.9\columnwidth]{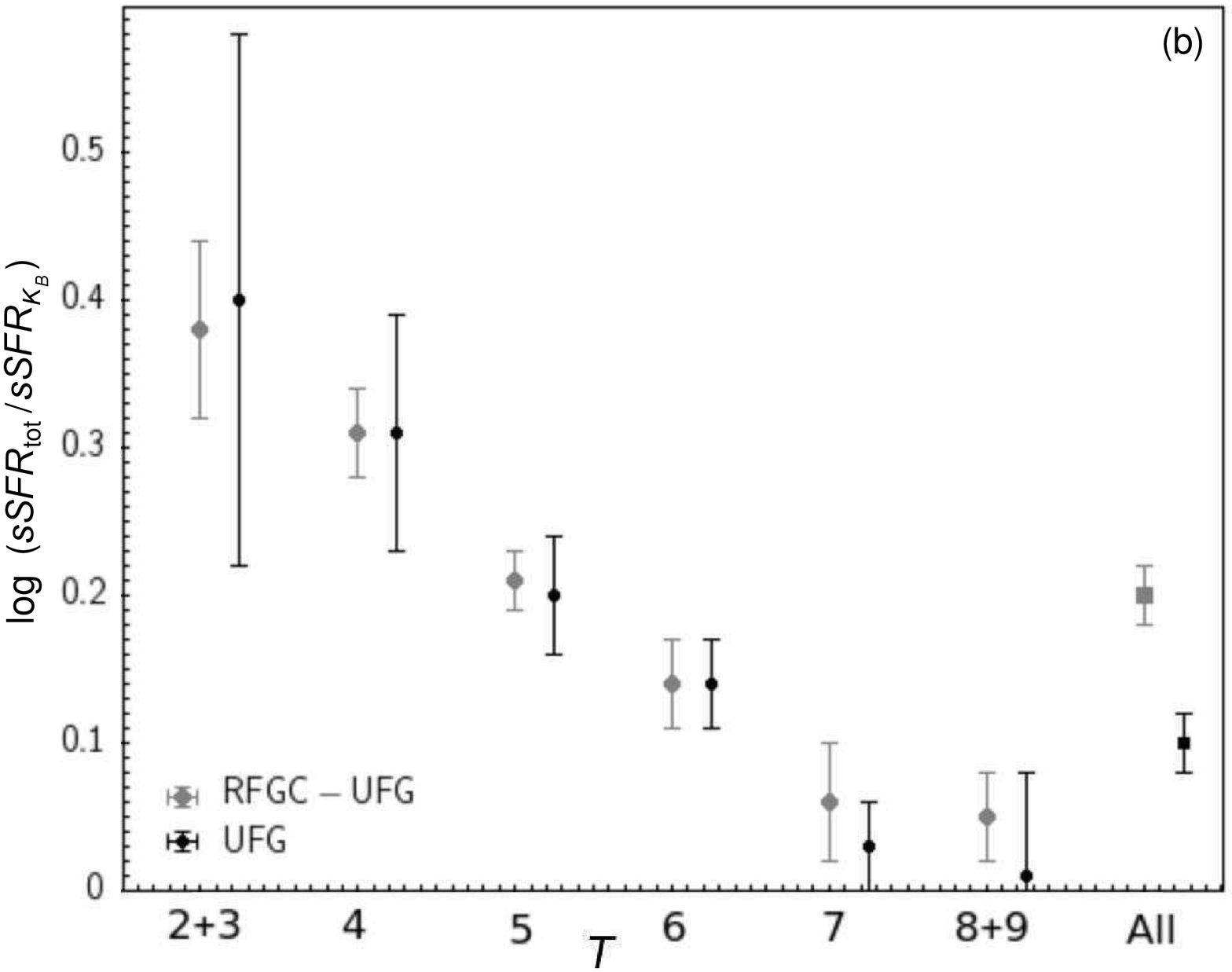}
\caption{Top: full (from FUV\,+\,IR) relative SFR for ultra-flat and flat galaxies as a function of
morphological type. Bottom: the ratio of specific SFR (full to the one determined from the \mbox{FUV-flux})
depending on the morphological type. The last pairs of points on
both panels are related to the averaged data for all types.}
\label{fig9:Melnyk_n_en}
\end{figure}

 \begin{figure}[]
\setcaptionmargin{5mm} \onelinecaptionstrue \captionstyle{normal}
\includegraphics[angle=0, width=0.9\columnwidth]{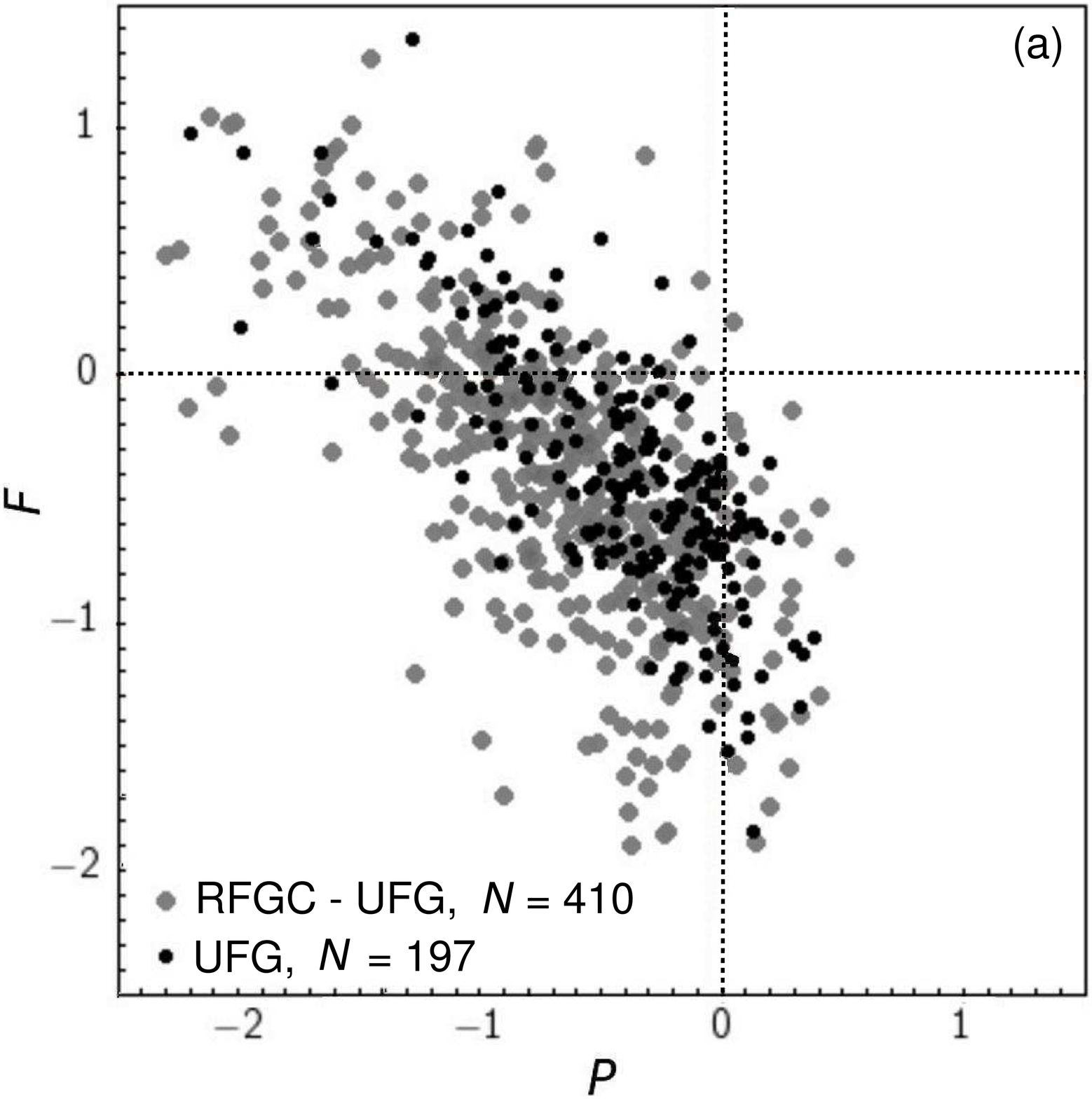}
\includegraphics[angle=0, width=0.9\columnwidth]{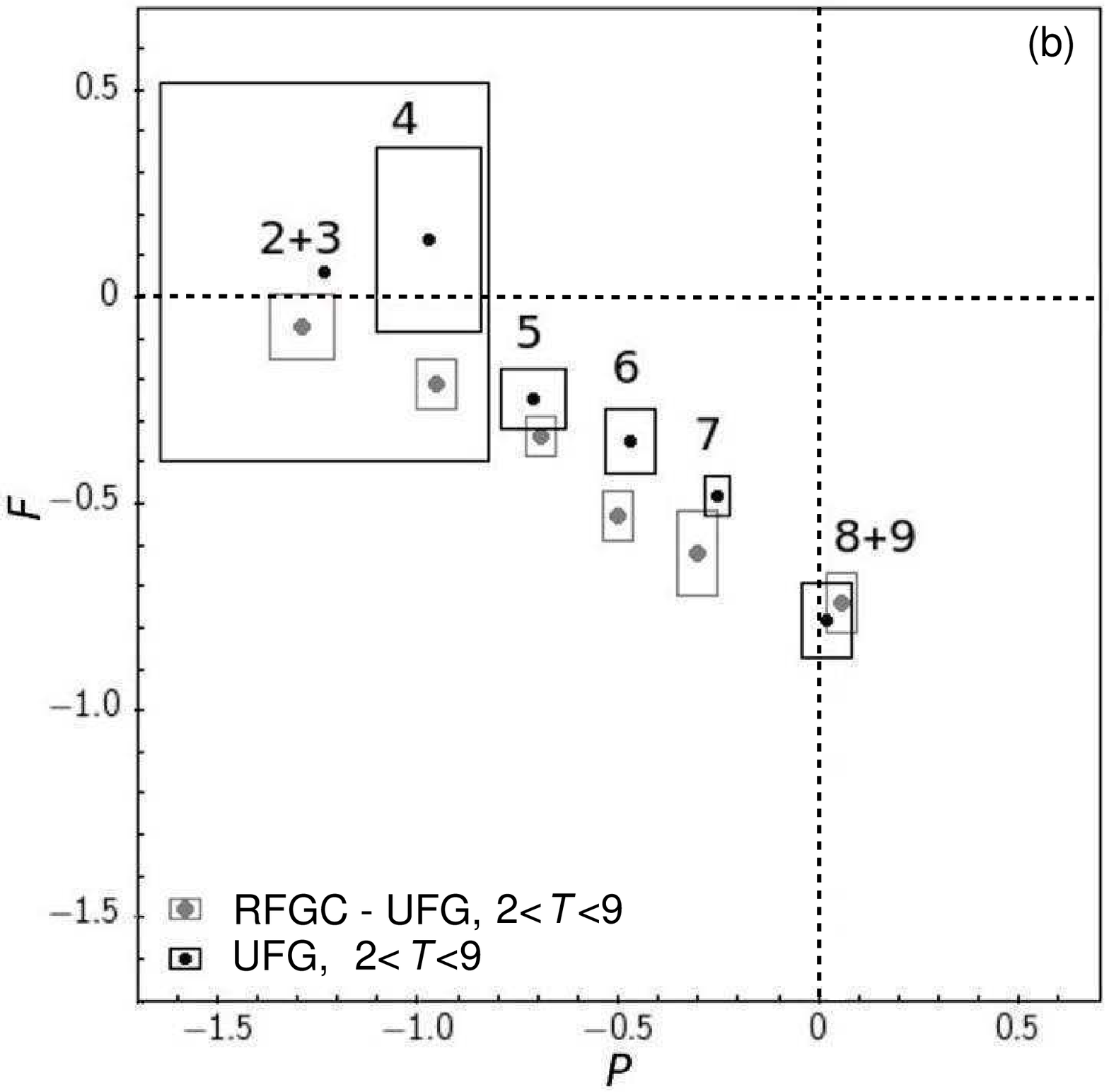}
\caption{The diagnostic ``Past--Future'' diagrams   for ultra-flat
and flat galaxies: (a)~the distribution of individual $P$ and   $F$
values; (b)~the mean $P$ and   $F$ for different morphological types.
The box sizes correspond to the errors in mean.}
\label{fig10:Melnyk_n_en}
 \end{figure}

Some authors (see~\cite{meu2016:Melnyk_n_en} and the literature cited
there) use not only the specific, but also the  effective star
formation rate, \mbox{ $SFRe = \log(SFR/M_{\rm H\,I})$}, to
characterize a galaxy. Here the hydrogen mass\linebreak \mbox{$M_{\rm
H\,I}=2.356 \times10^5\, D^2\, F_{\rm H\,I}$} is expressed in solar
mass units~\cite{hay1984:Melnyk_n_en}, $D$~is the distance in~Mpc,
and $F_{\rm H\,I}$ is the flux in Jy\,km\,s$^{-1}$, calculated
as\linebreak \mbox{$\log F_{\rm H\,I} = 0.4\,(17.4 - m_{21}$)}.
Figure~\ref{fig7:Melnyk_n_en} shows the dependence of the effective
SFR on the stellar  (the bottom panel) and hydrogen mass of galaxies.
Both diagrams for $SFRe$ look noticeably more diffuse than the
diagrams for $sSFR$. The distributions of ultra-flat and flat
galaxies do not show any particular mutual differences.

Integral flux of a galaxy in the  infrared band $W4~(22.0~\mu{\rm
m})$ is the \mbox{FUV-flux}-independent   SFR indicator. This is
especially important for ultra-flat galaxies due to  significant
internal extinction in the ultraviolet. According
to~\cite{jar2013:Melnyk_n_en}, the total star formation rate
$SFR_{\rm tot}$ is expressed by
\begin{equation}
SFR_{\rm tot} = SFR_{\rm FUV} + 0.83\,
SFR_{\rm IR}.
\end{equation}

The dependence of specific total star formation rate
 $\log\, sSFR_{\rm tot}$ on the stellar mass  for ultra-flat and flat
galaxies is shown in the diagram of Fig.~\ref{fig8:Melnyk_n_en}a.
Here the stellar mass was calculated from the $K$-luminosity
determined via relation~(9). Unlike Fig.~\ref{fig6:Melnyk_n_en}, this
panel shows all UFG  and RFGC--UFG galaxies with the  $SFR_{\rm tot}$
estimates. Figure~\ref{fig8:Melnyk_n_en}b shows the dependence of
$sSFR_{\rm tot}$ on the hydrogen-to-stellar-mass ratio of the galaxy.
The population of UFG galaxies on it looks much more compact than
that of the remaining RFGC galaxies.

The contribution of the infrared component in relation~(11) is on the
average small. For ultra-flat disks it amounts to $0.10\pm0.04$~dex,
and for \mbox{RFGC--UFG galaxies} this contribution is somewhat
larger: $0.20\pm0.03$~dex.
 As can be seen from the data
presented in Fig.~\ref{fig9:Melnyk_n_en}, the global SFR  in the
disks of flat galaxies, as well as the contribution to it of the IR
component  show a strong correlation with the morphological type. The
average value of  $\langle sSFR_{\rm tot} \rangle$ monotonically
increases by a magnitude from the   $T = 2,\,3$ types to  \mbox {$T =
8,\,9$} types. At that, the ratio of SFRs, determined by
equations~(11) and~(6), systematically drops from 0.4~dex for
\mbox{$T = 2+3$} to a virtually zero  excess for the
   \mbox{$T = 8+9$} types.
The differences of ultra-flat and flat galaxies in both panels of
Fig.~\ref{fig9:Melnyk_n_en} are within the random errors, but in
general the UFG sample has a higher SFR due to the presence in it of
a large fraction of late-type objects. Note that the account for the
IR component in  relation~(11) leaves all the galaxies below the
 $\log\,sSFR
=-9.4$ limit.

As noted by Karachentsev and Kaisin~\cite{kar2007:Melnyk_n_en}, the evolutionary
status of a galaxy is convenient to characterize  by dimensionless parameters
$P$ (Past) and  $F$ (Future),  which do not depend on the distance:
\begin{equation}
P = \log(SFR\times T_0/L_K) ,
\end{equation}
 \begin{equation}
F = \log[1.85\, M_{\rm H\,I}/(SFR\times T_0)] .
 \end{equation}
The $P$ parameter expresses the  specific star formation rate,
normalized to the age of the Universe \mbox{$T_0= 13.7$}~Gyrs.
Parameter $F$ (in the sense  opposite to the effective SFR) shows how
long the observed SFR can be maintained at the available gas reserves
in the disk.The coefficient 1.85 at $M_{\rm H\,I}$  takes into
account the contribution of helium and molecular hydrogen in the
overall mass of gas~\cite{fuk2004:Melnyk_n_en}.

Figure~\ref{fig10:Melnyk_n_en}   represents a diagnostic ``Past--Future'' diagram
for ultra-flat and flat galaxies. The top panel shows  individual
positions of each galaxy. The bottom panel demonstrates the mean
values of the  $P$ and   $F$ parameters and their errors for
different morphological types. As it can be seen, the overall
distribution of galaxies at the  $P$--$F$ diagram is diagonally
stretched, and disks of each morphological type are localized in a
certain part of the diagram.

\begin{turnpage}
\renewcommand\baselinestretch{0.55}
\begin{table*}
\setcaptionmargin{0mm} \onelinecaptionstrue \captionstyle{normal}
\caption{The mean values and the standard errors of the mean for the
calculated
 characteristics of flat and ultra-flat  galaxies depending on the
morphological type}
\medskip
\begin{tabular}{l|r@{$\,\pm\,$}l|r@{$\,\pm\,$}l|r@{$\,\pm\,$}l|r@{$\,\pm\,$}l|r@{$\,\pm\,$}l|r@{$\,\pm\,$}l|r@{$\,\pm\,$}l} \hline
  \multicolumn{1}{c|}{Parameter}  &\multicolumn{14}{c}{Morphological type}\\ \cline{2-15}
                & \multicolumn{2}{c|}{All types} & \multicolumn{2}{c|}{2\,+\,3} & \multicolumn{2}{c|}{4} & \multicolumn{2}{c|}{5} & \multicolumn{2}{c|}{6} & \multicolumn{2}{c|}{7} & \multicolumn{2}{c}{8\,+\,9} \\ \hline
$\log SFR$                & \multicolumn{2}{c|}{$N_{\rm RFGC-UFG}=531$}  & \multicolumn{2}{c|}{$45$} & \multicolumn{2}{c|}{$127$} & \multicolumn{2}{c|}{$152$} & \multicolumn{2}{c|}{$108$} & \multicolumn{2}{c|}{$45$}  & \multicolumn{2}{c}{$54$} \\
                          & $-0.59$ & $0.02$                             & $-0.73$ & $0.05$          & $-0.54$ & $0.04$           & $-0.45$ & $0.03$           & $-0.55$ & $0.05$           & $-0.75$ & $0.09$           & $-0.94$ & $0.09$  \\
                          & \multicolumn{2}{l|}{$N_{\rm UFG}=259$}       & \multicolumn{2}{c|}{$5$}  & \multicolumn{2}{c|}{$12$}  & \multicolumn{2}{c|}{$52$}  & \multicolumn{2}{c|}{$66$}  & \multicolumn{2}{c|}{$102$} & \multicolumn{2}{c}{$22$} \\
                          & $-0.45$ & $0.03$                             & $-0.7$ & $0.29$           & $-0.48$ & $0.11$           & $-0.47$ & $0.05$           & $-0.34$ & $0.05$           & $-0.42$ & $0.04$           & $-0.78$ & $0.14$  \\ \hline
$\log M_*$                & \multicolumn{2}{c|}{$N_{\rm RFGC-UFG}=722$}  & \multicolumn{2}{c|}{$67$} & \multicolumn{2}{c|}{$170$} & \multicolumn{2}{c|}{$201$} & \multicolumn{2}{c|}{$143$} & \multicolumn{2}{c|}{$67$}  & \multicolumn{2}{c}{$74$} \\
                          & $10.26$ & $0.02$                             & $10.76$ & $0.04$          & $10.6$ & $0.02$            & $10.4$ & $0.03$            & $10.1$ & $0.04$            & $9.81$ & $0.07$            & $9.33$ & $0.08$   \\
                          & \multicolumn{2}{l|}{$N_{\rm UFG}=333$}       & \multicolumn{2}{c|}{$8$}  & \multicolumn{2}{c|}{$16$}  & \multicolumn{2}{c|}{$70$}  & \multicolumn{2}{c|}{$80$}  & \multicolumn{2}{c|}{$134$} & \multicolumn{2}{c}{$25$}  \\
                          & $10.12$ & $0.03$                             & $10.59$ & $0.12$          & $10.62$ & $0.04$           & $10.36$ & $0.04$           & $10.24$ & $0.05$           & $9.97$ & $0.04$            & $9.32$ & $0.16$   \\ \hline
$\log M_{\rm H\,I}$       & \multicolumn{2}{c|}{$N_{\rm RFGC-UFG}=573$}  & \multicolumn{2}{c|}{$51$} & \multicolumn{2}{c|}{$132$} & \multicolumn{2}{c|}{$160$} & \multicolumn{2}{c|}{$120$} & \multicolumn{2}{c|}{$54$}  & \multicolumn{2}{c}{$56$}  \\
                          & $8.91$ & $0.03$                              & $9.09$ & $0.07$           & $9.16$ & $0.05$            & $9.09$ & $0.04$            & $8.77$ & $0.06$            & $8.61$ & $0.1$             & $8.29$ & $0.08$   \\
                          & \multicolumn{2}{l|}{$N_{\rm UFG}=258$}       & \multicolumn{2}{c|}{$7$}  & \multicolumn{2}{c|}{$12$}  & \multicolumn{2}{c|}{$55$}  & \multicolumn{2}{c|}{$62$}  & \multicolumn{2}{c|}{$102$} & \multicolumn{2}{c}{$20$}  \\
                          & $9.02$ & $0.03$                              & $9.01$ & $0.16$           & $9.5$ & $0.09$             & $9.15$ & $0.06$            & $9.11$ & $0.06$            & $8.98$ & $0.05$            & $8.41$ & $0.14$   \\ \hline
$\log (M_{\rm H\,I}/M_*)$ & \multicolumn{2}{c|}{$N_{\rm RFGC-UFG}=573$}  & \multicolumn{2}{c|}{$51$} & \multicolumn{2}{c|}{$132$} & \multicolumn{2}{c|}{$160$} & \multicolumn{2}{c|}{$120$} & \multicolumn{2}{c|}{$54$}  & \multicolumn{2}{c}{$56$}  \\
                          & $-1.34$ & $0.02$                             & $-1.66$ & $0.06$          & $-1.44$ & $0.04$           & $-1.31$ & $0.03$           & $-1.31$ & $0.05$           & $-1.19$ & $0.07$           & $-1.05$ & $0.07$  \\
                          & \multicolumn{2}{l|}{$N_{\rm UFG}=258$}       & \multicolumn{2}{c|}{$7$}  & \multicolumn{2}{c|}{$12$}  & \multicolumn{2}{c|}{$55$}  & \multicolumn{2}{c|}{$62$}  & \multicolumn{2}{c|}{$102$} & \multicolumn{2}{c}{$20$}  \\
                          & $-1.09$ & $0.02$                             & $-1.51$ & $0.1$           & $-1.11$ & $0.09$           & $-1.22$ & $0.04$           & $-1.08$ & $0.04$           & $-1.00$ & $0.03$           & $-1.06$ & $0.1$   \\ \hline
$\log sSFR$               & \multicolumn{2}{c|}{$N_{\rm RFGC-UFG}=531$}  & \multicolumn{2}{c|}{$45$} & \multicolumn{2}{c|}{$127$} & \multicolumn{2}{c|}{$152$} & \multicolumn{2}{c|}{$108$} & \multicolumn{2}{c|}{$45$}  & \multicolumn{2}{c}{$54$}  \\
                          & $-10.81$ & $0.02$                            & $-11.46$ & $0.06$         & $-11.13$ & $0.04$          & $-10.81$ & $0.03$          & $-10.64$ & $0.03$          & $-10.45$ & $0.05$          & $-10.14$ & $0.03$ \\
                          & \multicolumn{2}{l|}{$N_{\rm UFG}=259$}       & \multicolumn{2}{c|}{$5$}  & \multicolumn{2}{c|}{$12$}  & \multicolumn{2}{c|}{$52$}  & \multicolumn{2}{c|}{$66$}  & \multicolumn{2}{c|}{$102$} & \multicolumn{2}{c}{$22$}  \\
                          & $-10.56$ & $0.03$                            & $-11.37$ & $0.32$         & $-11.09$ & $0.11$          & $-10.83$ & $0.06$          & $-10.6$ & $0.04$           & $-10.38$ & $0.03$          & $-10.12$ & $0.05$ \\ \hline
$\log SFRe$               & \multicolumn{2}{c|}{$N_{\rm RFGC-UFG}=423$}  & \multicolumn{2}{c|}{$35$} & \multicolumn{2}{c|}{$97$}  & \multicolumn{2}{c|}{$122$} & \multicolumn{2}{c|}{$92$}  & \multicolumn{2}{c|}{$37$}  & \multicolumn{2}{c}{$40$}  \\
                          & $-9.48$ & $0.03$                             & $-9.8$ & $0.08$           & $-9.66$ & $0.06$           & $-9.53$ & $0.05$           & $-9.34$ & $0.06$           & $-9.25$ & $0.1$            & $-9.13$ & $0.07$  \\
                          & \multicolumn{2}{l|}{$N_{\rm UFG}=201$}       & \multicolumn{2}{c|}{$4$}  & \multicolumn{2}{c|}{$9$}   & \multicolumn{2}{c|}{$43$}  & \multicolumn{2}{c|}{$48$}  & \multicolumn{2}{c|}{$79$}  & \multicolumn{2}{c}{$18$}  \\
                          & $-9.48$ & $0.04$                             & $-9.93$ & $0.46$          & $-10.01$ & $0.22$          & $-9.62$ & $0.07$           & $-9.52$ & $0.08$           & $-9.39$ & $0.05$           & $-9.09$ & $0.09$  \\ \hline
$\log sSFR_t$             & \multicolumn{2}{c|}{$N_{\rm RFGC-UFG}=499$}  & \multicolumn{2}{c|}{$45$} & \multicolumn{2}{c|}{$124$} & \multicolumn{2}{c|}{$145$} & \multicolumn{2}{c|}{$100$} & \multicolumn{2}{c|}{$40$}  & \multicolumn{2}{c}{$45$}  \\
                          & $-10.61$ & $0.02$                            & $-11.08$ & $0.06$         & $-10.82$ & $0.03$          & $-10.6$ & $0.02$           & $-10.5$ & $0.03$           & $-10.39$ & $0.04$          & $-10.09$ & $0.03$ \\
                          & \multicolumn{2}{l|}{$N_{\rm UFG}=226$}       & \multicolumn{2}{c|}{$5$}  & \multicolumn{2}{c|}{$11$}  & \multicolumn{2}{c|}{$46$}  & \multicolumn{2}{c|}{$63$}  & \multicolumn{2}{c|}{$87$}  & \multicolumn{2}{c}{$14$}  \\
                          & $-10.46$ & $0.02$                            & $-10.97$ & $0.18$         & $-10.78$ & $0.08$          & $-10.63$ & $0.04$          & $-10.46$ & $0.03$          & $-10.35$ & $0.03$          & $-10.11$ & $0.07$ \\ \hline
$P$                       & \multicolumn{2}{c|}{$N_{\rm RFGC-UFG}=423$}  & \multicolumn{2}{c|}{$35$} & \multicolumn{2}{c|}{$97$}  & \multicolumn{2}{c|}{$122$} & \multicolumn{2}{c|}{$92$}  & \multicolumn{2}{c|}{$37$}  & \multicolumn{2}{c}{$40$}  \\
                          & $-0.65$ & $0.03$                             & $-1.29$ & $0.08$          & $-0.95$ & $0.05$           & $-0.69$ & $0.04$           & $-0.5$ & $0.04$            & $-0.3$ & $0.05$            & $0.06$ & $0.04$   \\
                          & \multicolumn{2}{l|}{$N_{\rm UFG}=201$}       & \multicolumn{2}{c|}{$4$}  & \multicolumn{2}{c|}{$9$}   & \multicolumn{2}{c|}{$43$}  & \multicolumn{2}{c|}{$48$}  & \multicolumn{2}{c|}{$79$}  & \multicolumn{2}{c}{$18$}  \\
                          & $-0.43$ & $0.03$                             & $-1.23$ & $0.41$          & $-0.97$ & $0.13$           & $-0.71$ & $0.08$           & $-0.47$ & $0.06$           & $-0.25$ & $0.03$           & $0.02$ & $0.06$   \\ \hline
$F$                       & \multicolumn{2}{c|}{$N_{\rm RFGC-UFG}=423$}  & \multicolumn{2}{c|}{$35$} & \multicolumn{2}{c|}{$97$}  & \multicolumn{2}{c|}{$122$} & \multicolumn{2}{c|}{$92$}  & \multicolumn{2}{c|}{$37$}  & \multicolumn{2}{c}{$40$}  \\
                          & $-0.39$ & $0.03$                             & $-0.07$ & $0.08$          & $-0.21$ & $0.06$           & $-0.34$ & $0.05$           & $-0.53$ & $0.06$           & $-0.62$ & $0.1$            & $-0.74$ & $0.07$  \\
                          & \multicolumn{2}{l|}{$N_{\rm UFG}=201$}       & \multicolumn{2}{c|}{$4$}  & \multicolumn{2}{c|}{$9$}   & \multicolumn{2}{c|}{$43$}  & \multicolumn{2}{c|}{$48$}  & \multicolumn{2}{c|}{$79$}  & \multicolumn{2}{c}{$18$}  \\
                          & $-0.39$ & $0.04$                             & $0.06$ & $0.46$           & $0.14$ & $0.22$            & $-0.25$ & $0.07$           & $-0.35$ & $0.08$           & $-0.48$ & $0.05$           & $-0.78$ & $0.09$  \\ \hline

\end{tabular}
\end{table*}
\renewcommand\baselinestretch{1.0}
\end{turnpage} 

The mean values of various parameters, characterizing the masses and
star formation rates in galaxies of different types are shown in
Table~2. Its structure is the same as that of Table~1. Generally the
differences of the mean parameters for the UFG  and RFGC\mbox{--}UFG
galaxies are small when compared within the same morphological type.
However, the resulting differences between the two considered samples
can be  significant because of the different proportion  in them of
the early and late types of galaxies. The difference of UFG  and
RFGC\mbox{--}UFG disks is most noticeable based on the relative
hydrogen abundance per   stellar mass unit.
 On the average, this difference amounts to  $0.25\pm0.03$~dex,
 somewhat varying from one morphological type to another.
 From this result,   distorted by the effects of observational selection in the least, it follows that the conversion of gas into stars in   very thin disks happens with a
noticeable delay compared with thicker disks.

\section{FINAL REMARKS}

The considered sample of ultra-flat galaxies is
a population of   spiral disks of the simplest structure, almost completely
deprived of the spheroidal component (bulge). In modern models of
galaxy formation it is assumed that such systems did not
experience  frequent mergers, residing in the areas of low number density
 of galaxies.  For this reason, ultra-flat galaxies can
serve as a reference sample for the study of the autonomous process
of star formation, where the tidal interactions of  neighbors are negligibly
small. At the same time, however, it is possible that the process of accretion of
intergalactic gas onto the thin disk has a significant impact on
the evolution of isolated  disks.

The diagnostic  ``Past--Future'' diagram (Fig.~\ref{fig10:Melnyk_n_en})
 is the most obvious way to compare the evolutionary
 status of galaxies from  various samples.
 As follows from the Table~2 data, the mean values of the $P$ and $F$ parameters  for the
ultra-flat galaxies amount to \mbox{$\langle P \rangle=
-0.43\pm0.03$} and \mbox {$\langle F \rangle = -0.39\pm0.04.$} For
the remaining flat RFGC--UFG galaxies these parameters are somewhat
different  from the previous ones: \mbox {$\langle P \rangle=
-0.65\pm0.03$} and   \mbox{$\langle F \rangle = -0.39\pm0.03$}.

Within the meaning of the  $P$ parameter,  a typical ultra-flat
galaxy is able to reproduce   only 1/3 of its present stellar mass at
the currently observed SFR. Consequently, the average SFR in thin
disks was in the past  approximately three times higher than in the
present. If we assume that the  $M_*/L_K$ ratio is not equal to
unity, but to  $0.5\,M_{\odot}/L_{\odot}$~\cite{mcg2015:Melnyk_n_en},
the average specific star formation rate in  ultra-flat disks in the
past would have been only by 0.13~dex, or  35\% higher than that
observed in the present. In this case, thin disks of galaxies would
look like regular factories uniformly processing gas into stars. At
that, the well-known observed   5--10-fold increase of the cosmic
$SFR(z)$ in the epoch  of \mbox{$z\sim
1$--$3$}~\cite{mad2014:Melnyk_n_en} would not have had any relation
with the evolution of thin disks of galaxies. The value of the
$\langle F \rangle$ parameter shows that a typical  ultra-flat galaxy
has reserves of gas, which allow to maintain the  observed SFR
 for nearly 6 more billion  years (if we do not
consider internal extinction of the H\,I flux therein).

As is seen in Fig.~\ref{fig10:Melnyk_n_en}, the $P$--$F$ diagram has an
elongation in the diagonal direction. This feature is
due to the systematic increase in the specific SFR along the Hubble sequence from the early to late types. The FUV-flux measurement errors
also lead to the scatter of
galaxies in the diagonal direction.

Considering the sample of isolated galaxies from the 2MIG
catalog~\cite{kara2010:Melnyk_n_en}, Melnyk et al.~\cite{mel2015:Melnyk_n_en} obtained the
mean parameters $\langle P \rangle= -0.62\pm0.02$ and\linebreak
\mbox{$\langle F \rangle = -0.14\pm0.02$}. The  $\langle  P \rangle$
value for \mbox {2MIG} galaxies coincides within the errors with the
mean\linebreak \mbox{$-0.65\pm0.03$} for flat galaxies. However, the
average parameter $\langle F \rangle=-0.39\pm0.03$
  for flat and ultra-flat
galaxies proves to be significantly smaller than the one for the
isolated 2MIG disks. The reason for this difference is obviously due
to  internal extinction of the H\,I flux in strongly inclined
galaxies. The RC3 catalog~\cite{vau1991:Melnyk_n_en} and HyperLEDA
give in addition to the \mbox{$m_{21}$-magnitude} the \mbox
{$m_{21}^c$-magnitudes} too, corrected for internal extinction,
according to~\cite{hei2000:Melnyk_n_en}:
\begin{eqnarray}
m^c_{21} & = & m_{21} - 2.5\, \log(0.031\, \sec i) \nonumber \\
         & + & 2.5 \log[1- exp(0.31\,\sec i)],
\end{eqnarray}
where $i$ is the angle of inclination. At $i> 89\degr$  the
extinction is considered in RC3 to be the same and equal to $0\fm82$.
 We did not use the $ m^c_{21}$-magnitude
for the \mbox{RFGC galaxies}, assuming that this scheme is not enough
applicable for the   edge-on disks: firstly, the errors in
determining the angle of inclination in them from the axial ratio
$a/b$ are sometimes considerable, and secondly, the amount of the
extinction has to depend not only on the angle $i$, but also on the
size (luminosity) of the galaxy. Nonetheless, the expected correction
for galaxies with $a/b>10$   based on the
scheme~\cite{hei2000:Melnyk_n_en}   is on the average $\Delta
m_{21}=0\fm5$, or $\Delta\log F_{\rm H\,I}=0.20$, the account of
which eliminates the differences in $\langle F \rangle$   between the
UFG  and 2MIG galaxies. Wherein the estimate of the characteristic
time of depletion of   gas in   flat galaxies increases from 6 to
9~Gyrs. Furthermore, for six flat galaxies, the kinematics of which
is investigated in~\cite{pet2016:Melnyk_n_en}, the mean
underestimation of the hydrogen mass due to the extinction in the
H\,I line amounts to $27\pm 6$\%.

For the LOG catalog of isolated galaxies of the Local
Supercluster~\cite{kar2011:Melnyk_n_en} the authors have found   median
parameters  $P=-0.05$ and $F=-0.03$. Comparing them with the values
of  $\langle P \rangle$ and $\langle F \rangle$ for the UFG sample,
it should be taken into account  that  dwarf galaxies with large gas
abundances and high SFR prevail in the near volume of the LOG
catalog. As we have already noted, the error of determination of the
FUV-flux in ultra-flat galaxies, residual uncertainty of the scheme
of corrections for the internal extinction, as well a  yet vague
systematics in estimating the stellar mass from the $K$-luminosity
altogether lead to the scatter of UFG galaxies by the  $P$ and   $F$
parameters  of about 0.2--0.3~dex.  This variation is somewhat
smaller than the one observed  (approximately by 0.4~dex). It is
possible that the true distribution of the thin isolated disks on the
diagnostic diagram $P$--$F$
 is very compact due to the uniform nature of the process
of conversion of gas into stars~\cite{oem2016:Melnyk_n_en}. To check
this assumption,   systematic programs of photometric and kinematic
studies of ultra-flat galaxies, including the measurements of their
H$\alpha$ fluxes  and determining the rotation curves from the
optical spectra are required.

\begin{acknowledgements}
In this paper, we used the    NED and HyperLEDA
 databases, as well as the data of  GALEX,
SDSS, 2MASS and WISE sky surveys. IDK thanks the Russian Science
Foundation for the financial support within the grant
no.~14-12-00965.
\end{acknowledgements}


\end{document}